\documentclass[twocolumn,epjc3]{svjour3}

\journalname{}
\def\makeheadbox{}

\usepackage{amsmath,graphicx}
\usepackage{multirow}
\usepackage{booktabs}
\usepackage{color}
\usepackage{soul}
\usepackage{comment}
\usepackage{txfonts}
\usepackage{cuted}
\usepackage{booktabs}
\usepackage{tabularx}

\definecolor{Green}{RGB}{255,0,0}
\definecolor{Red}{RGB}{255,0,0}

\RequirePackage[T1]{fontenc}

\smartqed  

\RequirePackage{graphicx}
\RequirePackage{mathptmx}      
\RequirePackage{flushend}
\RequirePackage[numbers,sort&compress]{natbib}
\RequirePackage[colorlinks,citecolor=blue,urlcolor=blue,linkcolor=blue]{hyperref}

\begin{document}
\title{Charged fluid tori with an approximate magnetic self-field around magnetized Schwarzschild black holes: A semi-analytic approach}
\author{Abylaikhan Tlemissov\thanksref{e1,addr1}
        \and
        Ji\v{r}\'{i} Kov\'{a}\v{r}\thanksref{e2,addr1} 
}
\thankstext{e1}{e-mail: tle0002@slu.cz}
\thankstext{e2}{e-mail: jiri.kovar@physics.slu.cz}
\institute{Research Centre for Theoretical Physics and Astrophysics, Institute of Physics, Silesian University in Opava,\\ Bezru\v{c}ovo n\'{a}m. 13, CZ-746\,01, Opava, Czech Republic\label{addr1}
}
\maketitle
\begin{abstract}
We study toroidal structures formed by a charged, non-self-gravitating perfect fluid orbiting a Schwarzschild black hole immersed in a poloidal magnetic field. To obtain analytic and semi-analytic solutions, we assume a cold, rest-mass-dominated fluid with uniform specific charge and zero conductivity, so the current is purely convective. This regime is formally opposite to the ideal GRMHD limit. The total magnetic field comprises a prescribed asymptotically uniform external component and an approximate self-field associated with the azimuthal current in the torus. We represent the latter by a regularized dipole approximation to the field of an effective circular current loop. Neglecting the self-field, the method of characteristics yields a general implicit solution of the integrability condition for a function parametrizing the specific-angular-momentum profile. A separately obtained cylindrically dependent class recovers a solution from our earlier work. For the radially dependent class, we integrate the Euler equations, obtain the enthalpy potential, and derive local conditions for equatorially centered tori. We then examine the field-free limit, both magnetic components separately, and their superposition. The external field supports closed tori within analytically derived parameter bounds, whereas no such equilibrium is found for the isolated approximate self-field within the numerically explored domain. In a representative combined-field sequence, the radial extent varies nonmonotonically with relative self-field strength: it increases for weak self-fields and decreases as the self-field becomes comparable to the external field. The results show how an approximate magnetic self-field modifies the morphology and domain of existence of charged-fluid tori.
\end{abstract}
\section{Introduction}

Black holes represent some of the most extreme environments in the
Universe, where strong gravitational fields profoundly influence the
behavior of matter and electromagnetic fields. In many astrophysical
systems, their immediate surroundings are filled with ionized plasma
whose charged constituents interact strongly with electromagnetic
fields. The interplay between gravity and electromagnetism is therefore
an essential ingredient of realistic models
\cite{Gammie2003,YuanNarayan2014}. Equilibrium configurations of charged
fluids around black holes are consequently of considerable theoretical
interest and may also be relevant to the interpretation of
observational phenomena associated with accreting compact objects.

The dynamics of charged matter in curved spacetime has been studied in
a variety of contexts, ranging from test-particle motion in combined
gravitational and electromagnetic fields to fully relativistic
magnetohydrodynamic simulations
\cite{AlievOzdemir2002,Font2008}. Although test-particle approaches
provide valuable insight into orbital properties and stability, they
neglect collective effects such as pressure gradients, spatial charge
distributions, and fluid interactions, which may become important in
sufficiently dense plasma environments \cite{Kulsrud2005}.

Charged-fluid structures may arise in accretion disks, coronae, and
magnetospheric environments in which departures from strict local
charge neutrality are permitted by strong electromagnetic fields. Such
structures have been modeled as electrically charged perfect fluids
orbiting magnetized black holes
\cite{Kovar2011,Sla-eta:2013,Kovar2014,Kovar2016,Tro-eta:2016,Tro-eta:2020,Dor-eta:2026}. A nonvanishing specific charge
introduces a Lorentz-force contribution into the relativistic Euler
equation and consequently modifies the equilibrium surfaces, the
locations of their centers, and the admissible toroidal morphologies.
These modifications may affect physical conditions relevant to
accretion variability, radiative processes, and the formation of
outflows \cite{BlandfordZnajek1977}.

Equilibrium analyses of relativistic fluids in black-hole spacetimes
provide a controlled framework for isolating the effects of gravity,
pressure, rotation, and electromagnetic interactions. Stationary and
axisymmetric perfect-fluid models based on the relativistic Euler
equation and effective-potential methods enable a systematic
investigation of how spacetime geometry and electromagnetic fields
determine the topology of equipotential surfaces and the morphology of
toroidal structures \cite{Abramowicz1978,Kozlowski1978}. Although
idealized, such analytic and semi-analytic constructions capture
essential properties of relativistic fluid equilibria and provide a
useful conceptual basis for interpreting more complex, fully
time dependent GRMHD models \cite{Komissarov1999}. Accordingly, rather
than relying exclusively on a numerical construction of the
equilibrium structures, we focus on analytic and semi-analytic methods
that make the integrability structure and the dependence on the model
parameters explicit.

The present model differs from standard ideal GRMHD primarily in its
treatment of electrical conductivity. Rather than adopting the
infinite-conductivity limit, we consider the zero-conductivity limit
and model the fluid as a single charged component without conductive
drift. Its current is therefore purely convective, and electric charge
is advected with the fluid. The electromagnetic field is not
constrained by the ideal-MHD condition. Consequently, although the
electrostatic self-field is neglected, the electric field measured in
the comoving frame need not vanish and can contribute directly to the
Lorentz-force balance.

We further adopt the cold-fluid approximation, in which the pressure
and internal-energy density are small compared with the rest-mass
density. Independently of this thermodynamic approximation, we assume
that the specific charge is spatially uniform. This choice differs from
those adopted in our previous studies
\cite{Kovar2016,2018PhRvD..97j4019T}, in which the specific-charge
distribution was allowed to vary throughout the fluid. The combination
of the cold-fluid approximation and a constant specific charge
considerably simplifies the equilibrium equations and enables a
transparent semi-analytic investigation of the admissible
angular-momentum profiles and toroidal structures.

In many stationary charged-fluid models, the electromagnetic field is
prescribed independently of the fluid. Here, we additionally include
an approximate representation of the magnetic field generated by the
azimuthal convective current within the torus. The total magnetic field
is represented as the superposition of a prescribed, asymptotically
uniform external field and an effective self-field modeled by a
regularized dipole approximation to the field of a circular current
loop. Both components are treated as test fields with respect to the
background spacetime, while their contributions to the Lorentz force
acting on the fluid are retained. This construction is
phenomenological: the effective loop current represents the distributed
azimuthal convective current within the torus, but its amplitude is
prescribed rather than determined self-consistently from Maxwell's
equations. We examine the two magnetic components separately and in
superposition, together with the field-free limit. This approach allows
us to investigate how the approximate magnetic self-field modifies the
equilibrium morphology and its domain of existence.

The paper is organized as follows. In
Section~\ref{R_euler_eq_sec}, we introduce the charged-fluid model,
specify its assumptions concerning the fluid motion, electromagnetic
field, and thermodynamic properties, and derive the integrability
condition for the relativistic Euler equations. Beginning with
Section~\ref{analytic_schwarzschild_sec}, we restrict our attention to
Schwarzschild spacetime and assume a spatially uniform specific charge
throughout the remainder of the paper. In contrast to our recent
studies, we prescribe the specific-charge profile and use the
integrability condition to determine the admissible
specific-angular-momentum profiles rather than prescribing the latter
in advance. In Section~\ref{analytic_schwarzschild_sec}, this condition
is expressed as a first-order quasilinear partial differential equation
for the angular-momentum parametrization function $u(r,\theta)$, whose
solutions permit the Euler equations to be integrated consistently. In
Section~\ref{uniform_field_solutions_sec}, we use the method of
characteristics to obtain a general implicit integral for an
asymptotically uniform external magnetic field with the magnetic
self-field neglected. We also recover separately the previously known
cylindrically dependent class $u=u(R)$
\cite{2020CQGra..37x5007K}. In
Section~\ref{radial_solutions_sec}, we investigate the radially
dependent class $u=u(r)$, integrate the Euler equations, and derive the
enthalpy potential and the local conditions for equatorially centered
toroidal structures. Within this class, we examine the external and
approximate self-field components separately and in superposition,
together with the field-free limit, and determine how the relative
self-field amplitude affects the radial extent of the resulting
toroidal structures.

Throughout the paper, we employ geometrized units with
$G=c=M_{\mathrm{BH}}=1$, where $M_{\mathrm{BH}}$ denotes the black-hole
mass, and use the metric signature $(-,+,+,+)$.
\section{Charged perfect fluid and the Euler equation}
\label{R_euler_eq_sec}
Neglecting viscosity and radiative processes, we model the fluid as a
charged perfect fluid. The associated total stress-energy tensor can
then be written as
\begin{equation}
    \mathcal{T}^{\mu\nu}
    =
    T^{\mu\nu}
    +
    T_{\mathrm{EM}}^{\mu\nu},
\end{equation}
where
\begin{align}
    T^{\mu\nu}
    &=
    (p+\epsilon)U^\mu U^\nu+p g^{\mu\nu},\\
    T_{\mathrm{EM}}^{\mu\nu}
    &=
    \frac{1}{4\pi}\left(\mathcal{F}^{\mu}{}_{\gamma}\mathcal{F}^{\nu\gamma}-\frac{1}{4}g^{\mu\nu}
        \mathcal{F}_{\gamma\delta}\mathcal{F}^{\gamma\delta}\right).
\end{align}
Here, $p$ denotes the thermal pressure of the fluid, excluding any
electromagnetic contribution; $\epsilon=\rho+e$ is its total energy
density, with $\rho$ and $e$ denoting the rest-mass and internal-energy
densities, respectively; and $U^\mu$ is the fluid four-velocity.

Note that throughout this work, we adopt the test-fluid and test-field
approximation. The charged fluid is affected by the fixed gravitational
background and by both the external and self-generated electromagnetic
fields, while the gravitational backreaction of the fluid and the
electromagnetic fields on the spacetime geometry is neglected.

The total electromagnetic field tensor is decomposed as
\begin{equation}
    \mathcal{F}^{\mu\nu}=F^{\mu\nu}+F_{\mathrm{self}}^{\mu\nu},
\end{equation}
where $F^{\mu\nu}$ and $F_{\mathrm{self}}^{\mu\nu}$ denote the
external- and self-field tensors, respectively. The external field is
assumed to be source-free in the region occupied by the fluid, whereas
the complete self-field is sourced by the fluid four-current:
\begin{equation}
    \nabla_\nu F^{\mu\nu}=0,
    \qquad
    \nabla_\nu F_{\mathrm{self}}^{\mu\nu}=4\pi J^\mu.
\end{equation}
The four-current density obeys the relativistic Ohm law
\begin{equation}
    J^\mu=q\rho U^\mu+\sigma\mathcal{F}^{\mu\nu}U_\nu,
\end{equation}
where $\sigma$ is the electrical conductivity and $q$ is the specific
charge of the fluid, which may in general depend on position.
The Maxwell equations above formally describe the complete
electromagnetic self-field. The magnetic-self-field approximation
adopted in the present study is specified below.

Using the conservation of the total stress-energy tensor,
$\nabla_\nu\mathcal{T}^{\mu\nu}=0$, together with Maxwell's equations,
we obtain the relativistic Euler equation in the form
\begin{equation}
    \partial_\mu p
    +
    U_\mu U^\nu\partial_\nu p
    +
    (p+\epsilon)U^\nu\nabla_\nu U_\mu
    =
    J^\nu\mathcal{F}_{\mu\nu}.
\end{equation}
The total electromagnetic field tensor is related to the total
four-potential $\mathcal{A}_\mu$ through
\begin{equation}
    \mathcal{F}_{\mu\nu}
    =
    \partial_\mu\mathcal{A}_\nu
    -
    \partial_\nu\mathcal{A}_\mu.
\end{equation}

\subsection{Rotation, conductivity, and field assumptions}

In what follows, we assume a stationary and axisymmetric spacetime and
stationary and axisymmetric electromagnetic fields in coordinates
$(t,r,\theta,\phi)$ adapted to these symmetries. We further assume
stationary and axisymmetric purely circular fluid motion,
\begin{equation}
    \partial_t U^\mu=\partial_\phi U^\mu=0,
    \qquad
    U^r=U^\theta=0.
\end{equation}

The fluid is modeled as a single charged component whose particles
share a common four-velocity. Since no conductive drift relative to the
fluid is allowed, the current is purely convective. We implement this
assumption by setting $\sigma=0$, so that the four-current density
reduces to
\begin{equation}
    J^\mu
    =
    q\rho U^\mu.
\end{equation}
Its only nonvanishing components are therefore
\begin{equation}
    J^t=q\rho U^t,
    \qquad
    J^\phi=q\rho U^\phi.
\end{equation}

Here, $J^t$ represents the charge-density contribution, whereas
$J^\phi$ is the azimuthal convective current that would source the
magnetic self-field in a fully self-consistent treatment. In the
present approximation, however, Maxwell's equations are not solved
self-consistently for this distributed current. Instead, its magnetic
field is represented phenomenologically by a regularized dipole
approximation to the field of an effective circular current loop with
prescribed amplitude $I$, while the electrostatic self-field associated
with $J^t$ is neglected.

We further assume that the external magnetic field and the retained
magnetic self-field are stationary, axisymmetric, and poloidal. Since
their electric components are neglected within the present
approximation, we set
\begin{equation}
    \mathcal{A}_t=0.
\end{equation}
For the poloidal magnetic fields, we choose a gauge in which
\begin{equation}
    \mathcal{A}_r=\mathcal{A}_\theta=0.
\end{equation}
Consequently, the only retained component of the total vector
potential can be decomposed as
\begin{equation}
    \mathcal{A}_\phi=A_\phi+A_\phi^{\mathrm{self}},
\label{Afi}
\end{equation}
where $A_\phi$ and $A_\phi^{\mathrm{self}}$ denote the external- and
self-field contributions, respectively.

Introducing the angular velocity and specific angular momentum,
\begin{equation}
    \Omega=\frac{U^\phi}{U^t},
    \qquad
    l=-\frac{U_\phi}{U_t},
    \label{Ol}
\end{equation}
the radial and polar components of the Euler equation reduce to
\begin{align}
    \frac{\partial_r p}{p+\epsilon}
    &=
    -\partial_r\ln|U_t|+\frac{\Omega\,\partial_r l}{1-\Omega l}
    +
    \frac{q\rho U^\phi}{p+\epsilon}\partial_r\mathcal{A}_\phi,
    \\
    \frac{\partial_\theta p}{p+\epsilon}
    &=
    -\partial_\theta\ln|U_t|+\frac{\Omega\,\partial_\theta l}{1-\Omega l}
    +
    \frac{q\rho U^\phi}{p+\epsilon}\partial_\theta\mathcal{A}_\phi.
\end{align}
\subsection{Thermodynamic assumptions}

To facilitate integration of the Euler equations, the thermodynamic
properties of the fluid must also be specified. Since our aim is to
perform an analytic integration, we assume a barotropic equation of
state given by the polytropic relation
\begin{equation}
    p=\kappa\rho^\Gamma,
\end{equation}
where $\kappa>0$ and $\Gamma>1$ are the polytropic constant and exponent,
respectively. We further assume that the fluid is isentropic. Thus, the
first law of thermodynamics, together with the polytropic equation of
state, gives
\begin{equation}
    \epsilon=\rho+e=\rho+\frac{p}{\Gamma-1}.
\end{equation}

The specific enthalpy is defined as
\begin{equation}
    h=\frac{\epsilon+p}{\rho}.
\end{equation}
For an isentropic fluid, the first law of thermodynamics implies
$\mathrm{d}p=\rho\,\mathrm{d}h$.
We then introduce the enthalpy potential $w$ through
\begin{equation}
    \mathrm{d}w
    =
    \frac{\mathrm{d}p}{p+\epsilon}.
\end{equation}
Since $p+\epsilon=\rho h$, it follows that
\begin{equation}
    \mathrm{d}w
    =
    \frac{\mathrm{d}h}{h}.
\end{equation}
Therefore, $w=\ln h+C$, where $C$ is an arbitrary additive constant. For the adopted
polytropic equation of state, $h\to1$ at the surface of the torus,
where $p\to0$. We normalize the potential by setting $w=0$ at the
surface. This corresponds to $C=0$, and hence
\begin{equation}
    w=\ln h.
\end{equation}
Its coordinate derivatives consequently satisfy
\begin{equation}
    \partial_\mu w
    =
    \frac{\partial_\mu p}{p+\epsilon}.
\end{equation}

Thus, in terms of $w$, the radial and polar components of the Euler
equation take the form
\begin{align}
    \partial_r w
    &=
    -\partial_r\ln|U_t|+\frac{\Omega\,\partial_r l}{1-\Omega l}
    +
    \frac{q\rho}{p+\epsilon}U^\phi\partial_r\mathcal{A}_\phi,
\label{euler_eq_r}
    \\
    \partial_\theta w
    &=
    -\partial_\theta\ln|U_t|+\frac{\Omega\,\partial_\theta l}{1-\Omega l}
    +
    \frac{q\rho}{p+\epsilon}U^\phi\partial_\theta\mathcal{A}_\phi.
\label{euler_eq_t}
\end{align}

In the cold-fluid limit, in which both the pressure and internal-energy
density are small compared with the rest-mass density,
$p\ll\rho$ and $e\ll\rho$, the total fluid energy density can be
approximated as $\epsilon=\rho+e\simeq\rho$. Consequently, $p+\epsilon\simeq\rho$, and the Euler equations reduce to
\begin{align}
    \partial_r w
    &=
    -\partial_r\ln|U_t|+\frac{\Omega\,\partial_r l}{1-\Omega l}
    +
    qU^\phi\partial_r\mathcal{A}_\phi
    \equiv\mathbb{R},
\label{euler_eq_1}
    \\
    \partial_\theta w
    &=-\partial_\theta\ln|U_t|+\frac{\Omega\,\partial_\theta l}{1-\Omega l}
    +
    qU^\phi\partial_\theta\mathcal{A}_\phi
    \equiv\mathbb{T}.
\label{euler_eq_2}
\end{align}
The solutions obtained from Eqs.~\eqref{euler_eq_1} and
\eqref{euler_eq_2} are physically consistent with the cold-fluid
approximation only where the specific enthalpy remains close to unity.
Indeed, using $w=\ln h$ and the adopted polytropic equation of state, we
have
\begin{equation}
    \frac{p}{\rho}
    =
    \frac{\Gamma-1}{\Gamma}\left(\mathrm{e}^w-1\right),
    \qquad
    \frac{e}{\rho}
    =
    \frac{1}{\Gamma}\left(\mathrm{e}^w-1\right).
\end{equation}
Thus, the assumptions $p\ll\rho$ and $e\ll\rho$ require
$0\leq w\ll1$ within the fluid.
\subsection{Integrability of the Euler equations and existence of tori}
\label{integrability_con_section}
For Eqs.~\eqref{euler_eq_1} and \eqref{euler_eq_2} to admit a common
local scalar solution $w(r,\theta)$, their right-hand sides must satisfy the
integrability condition
\begin{equation}
    \partial_\theta\mathbb{R}
    =
    \partial_r\mathbb{T}.
\label{IC}
\end{equation}

Using the definitions in Eq.~\eqref{Ol}, the angular velocity
$\Omega(r,\theta)$ is algebraically related to the specific angular
momentum $l(r,\theta)$ through
\begin{equation}
    \Omega
    =
    -\frac{g_{t\phi}+l g_{tt}}
           {g_{\phi\phi}+l g_{t\phi}}.
\end{equation}
Furthermore, the normalization condition $U^\mu U_\mu=-1$ yields
\begin{equation}
    U_t
    =
    -\sqrt{\frac{g_{t\phi}^{\,2}-g_{tt}g_{\phi\phi}}{g_{\phi\phi}+2l g_{t\phi}+l^2 g_{tt}}},
\end{equation}
while the azimuthal component of the four-velocity satisfies $U^\phi=\Omega U^t$.
Thus, for a prescribed background geometry $g_{\mu\nu}$, the quantities
$\Omega$, $U_t$, and $U^\phi$ are determined algebraically by the
specific-angular-momentum profile $l(r,\theta)$. Once the total
azimuthal electromagnetic potential $\mathcal{A}_\phi$ is also
specified, the integrability condition~\eqref{IC} becomes a partial
differential equation coupling the specific-angular-momentum and
specific-charge profiles, $l(r,\theta)$ and $q(r,\theta)$,
respectively. Consequently, these two profiles cannot be chosen
independently.

Once a mutually compatible pair of profiles $l(r,\theta)$ and
$q(r,\theta)$ has been found---with one profile prescribed and the other
determined from the integrability condition---the potential
$w(r,\theta)$ can be obtained by integrating the Euler equations. The
closed poloidal sections of its equipotential surfaces in the poloidal plane $(r,\theta)$
\begin{equation}
    w(r,\theta)=\mathrm{const},
\end{equation}
then determine the equilibrium toroidal structure of the rotating fluid. Moreover, since the equation of state is barotropic, surfaces of constant $w$ coincide with surfaces of constant
thermal pressure $p$ and rest-mass density $\rho$.

Thus, the center of the torus corresponds to a local maximum of $w$.
To locate its position, we first impose the necessary
stationarity conditions
\begin{equation}
    \partial_r w=0,
    \qquad
    \partial_\theta w=0.
\label{con_drw_general}
\end{equation}
Their simultaneous solutions determine the coordinates
$(r_{\rm c},\theta_{\rm c})$ of the candidate critical points. In the
remainder of the paper, we restrict our attention to structures
that are reflection-symmetric with respect to the equatorial plane and
whose candidate centers lie at $\theta_{\rm c}=\pi/2$. For such
structures, we define
\begin{equation}
    D_\theta
    \equiv
    \left.\partial_\theta w\right|_{(r_{\rm c},\pi/2)}=0,
\label{con_dthw}
\end{equation}
which holds identically by symmetry. The candidate center radii are
therefore determined by the remaining stationarity condition
\begin{equation}
    D_r
    \equiv
    \left.\partial_r w\right|_{(r_{\rm c},\pi/2)}=0.
\label{con_drw}
\end{equation}

A critical point is a strict nondegenerate local maximum if the
Hessian matrix of $w$ is negative definite. Equatorial reflection
symmetry also implies
\begin{equation}
    \left.\partial_{r\theta}w
    \right|_{(r_{\rm c},\pi/2)}
    =0.
\end{equation}
Consequently, the sufficient conditions for a strict nondegenerate
local maximum reduce to the radial and polar conditions
\begin{equation}
    D_{rr}
    \equiv
    \left.\partial_{rr}w \right|_{(r_{\rm c},\pi/2)}<0,
    \qquad
    D_{\theta\theta}
    \equiv
    \left.\partial_{\theta\theta}w\right|_{(r_{\rm c},\pi/2)}<0.
\label{con_drrththw}
\end{equation}

Furthermore, the boundary of a finite torus is determined by a closed
zero-equipotential surface, $w=0$. This surface intersects the
equatorial plane at the inner and outer edges, $r=r_{\rm in}$ and
$r=r_{\rm out}$. A limiting closed equipotential may develop a
self-crossing point at its inner edge,
$r_{\rm in}=r_{\rm cusp}$. This point is referred to as the cusp. If
the fluid fills or overfills the corresponding limiting surface,
fluid can escape through the cusp and accrete onto the black hole (see
Fig.~\ref{Fig:1} and also \cite{Tle-Kov:2024}).
\section{Charged-fluid tori with constant specific charge in Schwarzschild spacetime}
\label{analytic_schwarzschild_sec}

In what follows, we assume that the specific charge of the fluid is
spatially uniform, $q=\mathrm{const}$. We first seek
specific-angular-momentum profiles $l(r,\theta)$ satisfying the
integrability condition~\eqref{IC} and then determine the corresponding
enthalpy potentials $w(r,\theta)$ by integrating
Eqs.~\eqref{euler_eq_1} and \eqref{euler_eq_2}. Under the assumption of
constant specific charge, the integrability condition takes the
explicit form
\begin{equation}
\frac{\partial_\theta\Omega\,\partial_r l}{\left(1-l\Omega\right)^2}
+q\,\partial_\theta U^\phi\,\partial_r\mathcal{A}_\phi
=
\frac{\partial_r\Omega\,\partial_\theta l}{\left(1-l\Omega\right)^2}
+q\,\partial_r U^\phi\,\partial_\theta\mathcal{A}_\phi.
\label{equation_int_con1}
\end{equation}
Thus, for a prescribed value of $q$,
Eq.~\eqref{equation_int_con1} constitutes a first-order partial
differential equation for the specific-angular-momentum profile
$l(r,\theta)$.

Since our aim is to identify analytic solutions, we adopt Schwarzschild
spacetime as a simple background geometry, whose line element reads
\begin{equation}
\mathrm{d}s^2
=
-\left(1-\frac{2}{r}\right)\mathrm{d}t^2
+
\left(1-\frac{2}{r}\right)^{-1}\mathrm{d}r^2
+
r^2\mathrm{d}\theta^2
+
r^2\sin^2\theta\mathrm{d}\phi^2.
\label{Schwarzschild_metric}
\end{equation}
Even with this choice of background geometry, solving
Eq.~\eqref{equation_int_con1} directly remains cumbersome. We therefore
introduce a new angular-momentum parametrization function
$u(r,\theta)\neq0$, through which the sought
specific-angular-momentum profile is expressed in the convenient form
\begin{equation}
l(r,\theta)
=
\frac{r^{5/2}\sin^2\theta}{\sqrt{r-2}}\frac{1}{u(r,\theta)\sqrt{1+\left(\frac{r\sin\theta}{u(r,\theta)}
\right)^2}}.
\label{ang_momentum}
\end{equation}

We further assume that the total azimuthal component of the vector
potential associated with the poloidal magnetic field introduced in
Eq.~\eqref{Afi} can be written in the form
\begin{equation}
\mathcal{A}_{\phi}
=
A_{\phi}+A_{\phi}^{\mathrm{self}}
=
\frac{\mu}{2}r^2\sin^2\theta\,A(r),
\label{vector_potential_ansatz}
\end{equation}
where $\mu$ is a constant magnetic-field amplitude and $A(r)$ is a
radial profile specifying the magnetic configuration.

Substitution of Eqs.~\eqref{ang_momentum} and
\eqref{vector_potential_ansatz} into
Eq.~\eqref{equation_int_con1} yields
\begin{align}
&2r(r-2)\left(2+\bar{\mu}Au\right)\partial_r u\nonumber\\
&\quad
-\tan\theta\left[4(r-3)+\bar{\mu}(r-2)\left(2A+rA'\right)u\right]\partial_\theta u
=
4u,
\label{quasilinear_eq}
\end{align}
where we have introduced the effective electromagnetic parameter
\begin{equation}
\bar{\mu}=\mu q,
\label{effective_magnetic_parameter}
\end{equation}
and $A'=\partial_rA=\mathrm{d}A/\mathrm{d}r$.
Equation~\eqref{quasilinear_eq} is a first-order quasilinear partial
differential equation of the form
\begin{equation}
P(r,\theta,u)\partial_r u
+
Q(r,\theta,u)\partial_\theta u
=
T(r,\theta,u),
\end{equation}
with coefficients
\begin{align}
P(r,\theta,u)
&=2r(r-2)\left(2+\bar{\mu}Au\right),\\
Q(r,\theta,u)
&=
-\tan\theta\left[4(r-3)+\bar{\mu}(r-2)\left(2A+rA'\right)u\right],\\
T(r,\theta,u)
&=
4u.
\end{align}

Within the present ansatz, the function $A(r)$ encodes the radial
structure of the total magnetic field and can incorporate both a
prescribed external magnetic field and the magnetic self-field of the
torus, provided that both belong to the separable class of poloidal
fields defined by Eq.~\eqref{vector_potential_ansatz}.

In particular, the external component is represented by the
asymptotically uniform Wald test field, whose azimuthal vector
potential in Schwarzschild spacetime is
\begin{equation}
    A_{\phi}
    =
    \frac{B}{2}r^2\sin^2\theta,
    \label{uniform_vector_potential}
\end{equation}
where $B$ denotes the asymptotic magnetic-field strength
\cite{Wal:PRD-1974}. Exact general-relativistic solutions for the magnetic field generated
by a circular current loop around a Schwarzschild black hole were
derived in Ref.~\cite{Pet:PRD-1974}; see also
Refs.~\cite{Moss2011,KolosKofron2026}. In the present work, however,
we do not employ the exact current-loop solution. Instead, motivated
by its dipolar component, we represent the magnetic self-field of the
torus by the regularized profile
\begin{equation}
    A_{\phi}^{\mathrm{self}}
    =
    \frac{I r^2\sin^2\theta}{\left(r^2+r_0^2\right)^{3/2}},
\label{self_vector_potential}
\end{equation}
where $I$ is an effective magnetic-dipole amplitude related to the
azimuthal current carried by the torus. The distributed toroidal
current is modeled by an effective circular current loop lying in the
equatorial plane, while the regularization scale $r_0$ is identified
with its effective radius. This identification is phenomenological,
since Eq.~\eqref{self_vector_potential} represents a regularized
dipole approximation rather than the exact field of a circular current
loop. The profile remains regular in the central region and approaches
the standard dipolar form at large radii.

The total vector potential can therefore be written as
\begin{equation}
    \mathcal{A}_{\phi}
    =r^2\sin^2\theta\left(\frac{B}{2}+\frac{I}{\left(r^2+r_0^2\right)^{3/2}}\right).
\label{combined_vector_potential}
\end{equation}
Below, we consider the external and self-generated contributions
separately, in superposition, and in the field-free limit in which
both vanish. Their representations in terms of the general amplitude
$\mu$ and radial profile $A(r)$ are specified individually for each
configuration.

For each magnetic configuration, we first solve
Eq.~\eqref{quasilinear_eq} for the angular-momentum function $u$. Through Eq.~\eqref{ang_momentum}, the resulting solutions
determine the admissible specific-angular-momentum profiles
$l(r,\theta)$ consistent with the integrability
condition~\eqref{equation_int_con1}. Once this condition is satisfied,
we integrate the Euler equations~\eqref{euler_eq_1} and
\eqref{euler_eq_2} to obtain the corresponding enthalpy potential.
Equatorial toroidal equilibria are then identified by imposing the
center and local-maximum conditions~\eqref{con_drw} and
\eqref{con_drrththw}.

Regarding the functional dependence of $u$, we first investigate the
general class $u=u(R,z)$, by using the cylindrical coordinates $(R,z)$,
\begin{equation}
    R=r\sin\theta,\qquad z=r\cos\theta,\qquad r=\sqrt{R^2+z^2}, 
\end{equation}
including the particular case $u=u(R)$, for
an asymptotically uniform external magnetic field with the magnetic
self-field neglected. We then turn to the radially dependent class
$u=u(r)$, for which the analysis can be extended to configurations
including the regularized current-loop self-field, either alone or in
combination with the asymptotically uniform external field. A
comparably detailed analytic treatment of the current-loop self-field
for the general class $u=u(R,z)$ does not appear tractable within the
present framework and is therefore not pursued here.

For later use, we also express the magnetic field in the orthonormal frame
of a static observer located in the exterior Schwarzschild region. The
four-velocity of this observer is
\begin{equation}
n^\mu
=
\frac{1}{\sqrt{-g_{tt}}}\delta^\mu_t.
\end{equation}
For a purely poloidal magnetic field described by the covariant
azimuthal component $\mathcal{A}_{\phi}(r,\theta)$ of the vector
potential, the magnetic-field components measured by the static
observer are
\begin{equation}
B^{\hat{r}}=\frac{\partial_\theta\mathcal{A}_{\phi}
}{\sqrt{g_{\theta\theta}g_{\phi\phi}}},
\quad
B^{\hat{\theta}}=-\frac{\partial_r\mathcal{A}_{\phi}}{\sqrt{g_{rr}g_{\phi\phi}}},
\quad
B^{\hat{\phi}}=0.
\label{measured_magnetic_field}
\end{equation}
Here, hatted indices denote components measured in the observer's
local orthonormal frame. On the equatorial plane, the
$\hat{\theta}$ direction is normal to the plane and therefore represents
the vertical direction, up to its orientation.
\section{General class $u=u(R,z)$ in a uniform magnetic field}
\label{uniform_field_solutions_sec}
We now consider the simplest magnetic configuration that is a Schwarzschild
black hole immersed in an asymptotically uniform test magnetic field,
with the magnetic self-field neglected. For this configuration, we set
\begin{equation}
    A(r)=1,\qquad \mu=B,
\end{equation}
where $B$ denotes the magnetic-field strength. Consequently, the
effective electromagnetic parameter becomes $\bar{\mu}=\bar{B}=Bq$.

Equation~\eqref{quasilinear_eq} can then be conveniently rewritten in
cylindrical coordinates $(R,z)$.
For an arbitrary function $f(R,z)$, the corresponding derivatives
transform as
\begin{align}
    \partial_r f
    &=\frac{R}{r}\,\partial_R f+\frac{z}{r}\,\partial_z f,\\
    \partial_\theta f
    &=
    z\,\partial_R f-R\,\partial_z f.
\end{align}

Away from the equatorial plane, $z\neq0$,
Eq.~\eqref{quasilinear_eq} takes the form
\begin{equation}
    G(R,z,u)\,\partial_z u
    +R\,\partial_R u=u,
\label{quasilinear_eq_1}
\end{equation}
where
\begin{align}
    2zG(R,z,u)
    &=
    \bar{B}u\left(\sqrt{R^2+z^2}-2\right)\left(R^2+z^2\right)\nonumber\\
    &\quad
    +
    2R^2\left(\sqrt{R^2+z^2}-3\right)+2z^2\left(\sqrt{R^2+z^2}-2\right).
\label{G_uniform}
\end{align}
For reflection-symmetric solutions, this equation can be extended to
the equatorial plane by taking the appropriate limit from either
hemisphere.

Equation~\eqref{quasilinear_eq_1} can be solved using the method of
characteristics. The corresponding characteristic system is
\begin{equation}
    \frac{\mathrm{d}R}{\mathrm{d}s}=R,
    \qquad
    \frac{\mathrm{d}u}{\mathrm{d}s}=u,
    \qquad
    \frac{\mathrm{d}z}{\mathrm{d}s}=G(R,z,u),
    \label{characteristic_system_uniform}
\end{equation}
where $s$ parametrizes the characteristic curves.\footnote{Here and
henceforth, the upright symbol $\mathrm{e}$ denotes the base of the
natural logarithm and should not be confused with the italic symbol
$e$ used for the internal-energy density.}

The first two characteristic equations have the elementary solutions
\begin{equation}
    R=R_0\mathrm{e}^{s},
    \qquad
    u=u_0\mathrm{e}^{s}.
    \label{Ru_characteristics}
\end{equation}
It follows immediately that
\begin{equation}
    \frac{u}{R}
    =
    \frac{u_0}{R_0}
\end{equation}
is constant along each characteristic.

To integrate the third characteristic equation, we introduce
\begin{equation}
    t(s)
    =
    \sqrt{R(s)^2+z(s)^2}=r(s),
\end{equation}
so that
\begin{equation}
    z(s)
    =
    \pm\sqrt{t(s)^2-R_0^2\mathrm{e}^{2s}}.
\end{equation}
The third equation in
Eq.~\eqref{characteristic_system_uniform} then reduces to
\begin{equation}
    \frac{\mathrm{d}t}{\mathrm{d}s}
    =
    \left(1+\frac{\bar{B}u_0\mathrm{e}^{s}}{2}\right)(t-2)t.
\label{t_characteristic}
\end{equation}

Restricting our analysis to the region outside the event horizon,
$t=r>2$, integration of Eq.~\eqref{t_characteristic} gives
\begin{equation}
    \ln\left(1-\frac{2}{t}\right)
    =
    2s+\bar{B}u_0\mathrm{e}^{s}+c,
\end{equation}
where $c$ is an integration constant. Therefore,
\begin{equation}
    t(s)
    =
    \frac{2}{1-\mathrm{e}^{2s+\bar{B}u_0\mathrm{e}^{s}+c}}.
\end{equation}
Consequently,
\begin{equation}
    z(s)
    =
    \pm\sqrt{\frac{4}{\left(1-\mathrm{e}^{2s+\bar{B}u_0\mathrm{e}^{s}+c}\right)^2}
        -R_0^2\mathrm{e}^{2s}}.
\label{z_characteristic}
\end{equation}
On the exterior branch, the characteristic parameters must satisfy
\begin{equation}
    2s+\bar{B}u_0\mathrm{e}^{s}+c<0,
\end{equation}
which ensures that $t(s)>2$. In addition, the radicand in
Eq.~\eqref{z_characteristic} must be nonnegative.
\subsection{General integral}

The general integral follows from the two independent characteristic
invariants,
\begin{equation}
    \frac{u}{R},
    \quad
    \frac{1}{R^2}\left(1-\frac{2}{\sqrt{R^2+z^2}}\right)\mathrm{e}^{-\bar{B}u}.
\end{equation}
It can be written implicitly as
\begin{equation}
    F\left(\frac{u}{R}\right)
    \mathrm{e}^{\bar{B}u}
    =
    \frac{1}{R^2}\left(1-\frac{2}{\sqrt{R^2+z^2}}\right),
\label{gen_sol_uniform_2}
\end{equation}
where $F$ is an arbitrary function. Using
$R^2=g_{\phi\phi}$ and
$1-2/r=-g_{tt}$ in Schwarzschild spacetime,
Eq.~\eqref{gen_sol_uniform_2} can equivalently be expressed as
\begin{equation}
    F\left(\frac{u}{\sqrt{g_{\phi\phi}}}
    \right)\mathrm{e}^{\bar{B}u}
    =
    -\frac{g_{tt}}{g_{\phi\phi}}.
\label{F_equation}
\end{equation}
This general family depends explicitly on the effective magnetic-field
parameter $\bar{B}=Bq$.
\subsection{Cylindrically dependent class $u=u(R)$}

A particular family of solutions is obtained separately by assuming
that the angular-momentum parametrization function depends only on the
cylindrical radius, $u=u(R)$. In this case,
Eq.~\eqref{quasilinear_eq_1} reduces to
\begin{equation}
    R\frac{\mathrm{d}u}{\mathrm{d}R}=u,
\end{equation}
and hence $u/R$ is constant. Restricting attention to the chosen sense
of rotation and parametrizing the positive constant by $0<k<1$, we
obtain
\begin{equation}
    u
    =
    \sqrt{\frac{1-k}{k}}\,R.
\label{gen_sol_uniform_1}
\end{equation}
According to Eq.~\eqref{ang_momentum}, this solution corresponds to the
specific-angular-momentum distribution
\begin{equation}
    l
    =
    \left(-k\frac{g_{\phi\phi}}{g_{tt}}\right)^{1/2}.
\label{CQG_sol}
\end{equation}
Using $-g_{tt}/g_{\phi\phi}=\Omega/l$, this distribution satisfies
\begin{equation}
    \Omega l=k, \qquad \Omega=\frac{k}{l}.
\end{equation}
It therefore provides an explicit nonconstant rotation law of the
form $\Omega=\Omega(l)$.

In contrast to the general family, this particular distribution is
independent of the effective magnetic-field parameter $\bar{B}$. The
corresponding solution of Eqs.~\eqref{euler_eq_1} and
\eqref{euler_eq_2} for the angular-momentum
profile~\eqref{CQG_sol} was presented in
Ref.~\cite{2020CQGra..37x5007K}, together with a representative
toroidal structure.
\subsection{Field-free and uncharged-fluid limits}
In the limiting cases in which the Lorentz-force contribution to the
Euler equations vanishes, namely, the field-free limit $B=0$ and the
uncharged-fluid limit $q=0$, the effective electromagnetic parameter
also vanishes, $\bar{B}=Bq=0$. Consequently,
Eq.~\eqref{F_equation} reduces to
\begin{equation}
    F\left(\frac{u}{\sqrt{g_{\phi\phi}}}\right)
    =
    -\frac{g_{tt}}{g_{\phi\phi}}.
\end{equation}
Assuming that $F$ is locally invertible and denoting its inverse by an
arbitrary function $\tilde{F}$, we obtain
\begin{equation}
    u
    =
    \sqrt{g_{\phi\phi}}\,\tilde{F}\left(-\frac{g_{tt}}{g_{\phi\phi}}\right).
\end{equation}
Using Eq.~\eqref{ang_momentum}, the corresponding
specific-angular-momentum distribution can be written as
\begin{equation}
    l
    =
    \left(-\frac{g_{\phi\phi}}{g_{tt}}\right)^{1/2}\frac{1}{\sqrt{1+\tilde{F}\left(-\frac{g_{tt}}{g_{\phi\phi}}\right)^2}}
    \equiv
    \mathcal{G}\left(-\frac{g_{tt}}{g_{\phi\phi}}\right).
\label{ellF}
\end{equation}

Since, in a static spacetime,
$-g_{tt}/g_{\phi\phi}=\Omega/l$, Eq.~\eqref{ellF} is locally
equivalent, for nonconstant specific angular momentum, to the familiar
integrability condition $\Omega=\Omega(l)$ for a fluid subject to no
Lorentz force. Constant-$l$ profiles form a degenerate, automatically
integrable subclass. Indeed, for $\bar{B}=0$,
Eqs.~\eqref{euler_eq_r} and \eqref{euler_eq_t} reduce to
\begin{align}
    \partial_r w
    &=-\partial_r\ln|U_t|+\frac{\Omega\,\partial_r l}{1-\Omega l},
    \\
    \partial_\theta w
    &=-\partial_\theta\ln|U_t|+\frac{\Omega\,\partial_\theta l}{1-\Omega l}.
\end{align}
The integrability properties of these equations were discussed in
detail in the seminal papers
\cite{Abramowicz1978,Kozlowski1978}.

\section{Radially dependent class $u=u(r)$}
\label{radial_solutions_sec}

Since $A=A(r)$ depends only on the radial coordinate, it is natural to
consider the particular class of solutions for which $u=u(r)$. Under
this assumption, the integrability condition~\eqref{quasilinear_eq}
reduces to the ordinary differential equation
\begin{equation}
    r(r-2)\left(2+\bar{\mu}Au\right)u'=2u,
\label{radial_equation}
\end{equation}
where $u'=\partial_r u=\mathrm{d}u/\mathrm{d}r$. For
$\bar{\mu}\neq0$, we introduce the rescaled parametrization function
\begin{equation}
    \overline{u}
    =
    \bar{\mu}u.
\end{equation}
Since $\bar{\mu}$ is constant, Eq.~\eqref{radial_equation} becomes
\begin{equation}
    r(r-2)\left(2+A\overline{u}\right)\overline{u}'=2\overline{u},
\label{radial_equation_bar}
\end{equation}
where
$\overline{u}'=\mathrm{d}\overline{u}/\mathrm{d}r$.
The limiting case $\bar{\mu}=0$ will be treated separately below using
the unscaled function $u$.

In terms of $\overline{u}$, the radial and polar components of the
Euler equation take the form
\begin{align}
    \partial_r w
    &=
    -\frac{1}{r(r-2)}\nonumber\\
    &\quad
    +\frac{\bar{\mu}^{\,2}r}{2(r-2)\overline{u}^{\,2}}\left[2(r-3)+(r-2)\overline{u}
        \left(2A+rA'\right)\right]\sin^2\theta,
\label{drw}
    \\
    \partial_\theta w
    &=
    \frac{\bar{\mu}^{\,2}r^2}{\overline{u}^{\,2}}\left(1+A\overline{u}\right)\sin\theta\cos\theta.
\label{dthw}
\end{align}

Integrating Eq.~\eqref{dthw} with respect to $\theta$ gives
\begin{equation}
    w(r,\theta)
    =
    \frac{\bar{\mu}^{\,2}r^2}{2\overline{u}^{\,2}}\left(1+A\overline{u}\right)\sin^2\theta+v(r),
\label{w_intermediate}
\end{equation}
where $v(r)$ is an arbitrary function of the radial coordinate.
Differentiating Eq.~\eqref{w_intermediate} with respect to $r$ and
comparing the result with Eq.~\eqref{drw}, we obtain
\begin{align}
    v'(r)
    &=-\frac{1}{r(r-2)}\nonumber\\
    &\quad
    +\frac{\bar{\mu}^{\,2}r}{2(r-2)\overline{u}^{\,3}}\left[-2\overline{u}+r(r-2)
        \left(2+A\overline{u}\right)\overline{u}'\right]\sin^2\theta.
\label{v_derivative_general}
\end{align}
Consistency with the assumption $v=v(r)$ requires the coefficient of
$\sin^2\theta$ to vanish. This requirement is satisfied identically by
Eq.~\eqref{radial_equation_bar}. Consequently,
\begin{equation}
    v'(r)
    =
    -\frac{1}{r(r-2)}.
\end{equation}
Integration then gives
\begin{equation}
    v(r)
    =
    -\frac{1}{2}\ln\left(1-\frac{2}{r}\right)+v_{\mathrm{in}},
\end{equation}
where $v_{\mathrm{in}}$ is an integration constant. This constant will
be fixed below by requiring the enthalpy potential to vanish at the
inner edge of the torus,
$w(r_{\mathrm{in}},\pi/2)=0$.

The corresponding enthalpy potential is therefore
\begin{align}
    w(r,\theta)
    &=
    -\frac{1}{2}\ln\left(1-\frac{2}{r}\right)\nonumber\\
    &\quad
    +\frac{\bar{\mu}^{\,2}r^2}{2\overline{u}^{\,2}}\left(1+A\overline{u}\right)
    \sin^2\theta+v_{\mathrm{in}}.
\label{w_function}
\end{align}
After the rescaling $\overline{u}=\bar{\mu}u$,
Eq.~\eqref{radial_equation_bar} is independent of $\bar{\mu}$, whereas
Eq.~\eqref{w_function} depends on $\bar{\mu}$ only through
$\bar{\mu}^{\,2}$. Hence, the family of enthalpy potential solutions is
invariant under the simultaneous transformation $\bar{\mu}\rightarrow-\bar{\mu}$, $u\rightarrow-u$,
which leaves $\overline{u}=\bar{\mu}u$ unchanged.

A complete investigation of closed toroidal structures requires
solving Eq.~\eqref{radial_equation_bar} for a specified radial profile
$A(r)$. Nevertheless, the equatorial stationarity condition and the
polar-maximum condition can already be evaluated without an explicit
solution. For brevity, we denote the value of the rescaled
parametrization function at a candidate center by $\overline{u}_{\rm c}\equiv
    \overline{u}(r_{\rm c})$.

Equation~\eqref{dthw} shows directly that the stationarity condition
$D_\theta=0$ is satisfied identically at $\theta=\pi/2$, consistently
with equatorial reflection symmetry. Moreover, differentiating
Eq.~\eqref{dthw} with respect to $\theta$ and evaluating the result at
the equatorial candidate center gives
\begin{equation}
    D_{\theta\theta}
    =
    -\frac{\bar{\mu}^{\,2}r_{\rm c}^{\,2}}{\overline{u}_{\rm c}^{\,2}}\left[1+A(r_{\rm c})\overline{u}_{\rm c}\right].
\end{equation}
It therefore follows directly from Eq.~\eqref{dthw} that the polar
condition $D_{\theta\theta}<0$ is equivalent to
\begin{equation}
    1+
    A(r_{\rm c})\overline{u}_{\rm c}
    >0,
    \label{d2wtheta}
\end{equation}
provided that $\bar{\mu}\neq0$ and
$\overline{u}_{\rm c}\neq0$.

The remaining stationarity condition $D_r=0$, evaluated using
Eq.~\eqref{drw}, yields
\begin{equation}
    \bar{\mu}^{\,2}
    = 
    \frac{\overline{u}_{\rm c}^{\,2}}{r_{\rm c}^{\,2}\left\{r_{\rm c}-3+(r_{\rm c}-2)\left[A(r_{\rm c})+\frac{r_{\rm c}}{2}A'(r_{\rm c})\right]\overline{u}_{\rm c}\right\}}.
    \label{center_parameter_relation}
\end{equation}
Provided that $r_{\rm c}>2$ and
$\overline{u}_{\rm c}\neq0$,
Eq.~\eqref{center_parameter_relation} yields a real nonzero value of
$\bar{\mu}$ if and only if
\begin{equation}
    r_{\rm c}-3+(r_{\rm c}-2)\left[A(r_{\rm c})+\frac{r_{\rm c}}{2}A'(r_{\rm c})\right]\overline{u}_{\rm c}>0.
\label{center_admissibility_condition}
\end{equation}

The radial-maximum condition $D_{rr}<0$ still depends on the specified
magnetic profile $A(r)$. For later specialization, however, it is
useful to derive its general form directly from Eq.~\eqref{drw}. We
introduce the center-evaluated quantities
\begin{equation}
    A_{\rm c}\equiv A(r_{\rm c}),
    \qquad
    A'_{\rm c}\equiv A'(r_{\rm c}),
    \qquad
    A''_{\rm c}\equiv A''(r_{\rm c}),
\end{equation}
together with
\begin{equation}
    \overline{u}'_{\rm c}\equiv\overline{u}'(r_{\rm c})
    =
    \frac{2\overline{u}_{\rm c}}{r_{\rm c}(r_{\rm c}-2)\left(2+A_{\rm c}\overline{u}_{\rm c}\right)},
\label{ubar_derivative_center}
\end{equation}
where Eq.~\eqref{radial_equation_bar} has been used.

For compactness, we further define
\begin{equation}
    \mathcal{K}_{\rm c}
    \equiv
    2(r_{\rm c}-3)+(r_{\rm c}-2)\overline{u}_{\rm c}\left(2A_{\rm c}
        +r_{\rm c}A'_{\rm c}\right),
\label{K_center}
\end{equation}
and
\begin{align}
    \mathcal{K}'_{\rm c}
    &\equiv
    2+\overline{u}_{\rm c}\left(2A_{\rm c}+r_{\rm c}A'_{\rm c}\right)\nonumber\\
    &\quad
    +(r_{\rm c}-2)\overline{u}'_{\rm c}\left(2A_{\rm c}+r_{\rm c}A'_{\rm c}\right)\nonumber\\
    &\quad
    +(r_{\rm c}-2)\overline{u}_{\rm c}\left(3A'_{\rm c}+r_{\rm c}A''_{\rm c}\right).
\label{K_derivative_center}
\end{align}
Differentiating Eq.~\eqref{drw} with respect to $r$ and evaluating the
result at the equatorial candidate center then gives
\begin{align}
    D_{rr}
    &=
    \frac{2(r_{\rm c}-1)}{r_{\rm c}^{\,2}(r_{\rm c}-2)^2}\nonumber\\
    &\quad
    +\frac{\bar{\mu}^{\,2}}{(r_{\rm c}-2)\overline{u}_{\rm c}^{\,2}}
    \left[\frac{r_{\rm c}\mathcal{K}'_{\rm c}}{2}-\frac{\mathcal{K}_{\rm c}}{(r_{\rm c}-2)}
        -\frac{r_{\rm c}\mathcal{K}_{\rm c}\overline{u}'_{\rm c}}{\overline{u}_{\rm c}}\right].
\label{Drr_general}
\end{align}
Together with the stationarity relation
Eq.~\eqref{center_parameter_relation},
Eq.~\eqref{Drr_general} determines the radial-maximum condition for
any specified twice-differentiable magnetic profile $A(r)$.

\subsection{Field-free and uncharged-fluid limits}

In the limiting cases in which the Lorentz-force contribution to the
Euler equations vanishes---namely, the field-free limit, realized by
$A(r)\equiv0$ or $\mu=0$, and the uncharged-fluid limit $q=0$---all
electromagnetic terms drop out of the equilibrium equations. These
equations consequently reduce to the same form as those of an
uncharged perfect fluid in Schwarzschild spacetime. Although
constant-$l$ tori are known to exist in this setting
\cite{Abramowicz1978,Kozlowski1978}, we restrict our analysis to the
radially dependent class $u=u(r)\neq0$ defined by
Eq.~\eqref{ang_momentum}.

In any of these limiting cases, Eq.~\eqref{radial_equation} becomes
\begin{equation}
    r(r-2)u'=u,
\end{equation}
with the solution
\begin{equation}
    u(r)=u_{\infty}\sqrt{1-\frac{2}{r}}=u_{\infty}\sqrt{-g_{tt}},
\label{u_zero_field}
\end{equation}
where $u_{\infty}\neq0$ is the asymptotic value of $u$. The
corresponding nonconstant specific-angular-momentum profile can be
written as
\begin{equation}
    l=\frac{1}{\Omega}-u_{\infty}^{\,2}\Omega.
\label{l_zero_field}
\end{equation}

Before integrating the Euler equations, we note that
Eqs.~\eqref{drw} and \eqref{dthw} are written in terms of the rescaled
function $\overline{u}=\bar{\mu}u$. Therefore, the limit
$\bar{\mu}\to0$ cannot be taken by simply discarding the terms
containing the explicit factor $\bar{\mu}^{\,2}$, because
\begin{equation}
    \frac{\bar{\mu}^{\,2}}{\overline{u}^{\,2}}
    =
    \frac{1}{u^2}
\end{equation}
remains finite. Rewriting the Euler equations in terms of the unscaled
function $u$ and then taking any of the interaction-free limits
considered here gives
\begin{align}
    \partial_r w
    &=-\frac{1}{r(r-2)}+\frac{r(r-3)}{(r-2)u^2}\sin^2\theta,
\label{drw_zero_field}
    \\
    \partial_\theta w
    &=
    \frac{r^2}{u^2}\sin\theta\cos\theta.
\label{dthw_zero_field}
\end{align}
The terms proportional to the electromagnetic interaction have
vanished, whereas the rotational contributions remain. Integrating
Eqs.~\eqref{drw_zero_field} and \eqref{dthw_zero_field} and using
Eq.~\eqref{u_zero_field}, we obtain
\begin{equation}
    w(r,\theta)
    =
    -\frac{1}{2}\ln\left(1-\frac{2}{r}\right)+\frac{r^3}{2u_{\infty}^{\,2}(r-2)}\sin^2\theta+v_{\mathrm{in}}.
\label{w_zero_field}
\end{equation}

The polar-maximum condition is satisfied automatically, since
\begin{equation}
    D_{\theta\theta}=-\frac{r_{\rm c}^{\,3}}{u_{\infty}^{\,2}(r_{\rm c}-2)}<0
\end{equation}
for $r_{\rm c}>2$. The radial stationarity condition $D_r=0$ yields
\begin{equation}
    u_{\infty}^{\,2}
    =
    \frac{r_{\rm c}^{\,3}(r_{\rm c}-3)}{r_{\rm c}-2},
\label{zero_field_center_condition}
\end{equation}
which requires $r_{\rm c}>3$. However, at every candidate center
radius satisfying Eq.~\eqref{zero_field_center_condition}, the radial
second derivative is
\begin{equation}
    D_{rr}=\frac{(3r_{\rm c}-7)^2+5}{3r_{\rm c}^{\,2}(r_{\rm c}-2)^2(r_{\rm c}-3)}>0.
\label{zero_field_radial_second_derivative}
\end{equation}
Thus, these candidate center radii correspond to radial minima rather
than radial maxima of $w$, and no equatorially centered closed
toroidal equilibrium exists within the radially dependent class
$u=u(r)$. Note that this conclusion does not apply to the standard
constant-$l$ tori, which lie outside this restricted class.
\subsection{Uniform magnetic field}
We now return to the asymptotically uniform test magnetic-field
configuration considered in
Section~\ref{uniform_field_solutions_sec}, for which we set
\begin{equation}
    A(r)=1,
    \qquad
    \mu=B,
\end{equation}
where $B$ denotes the magnetic-field strength. Consequently, the
effective electromagnetic parameter is again $\bar{\mu}=\bar{B}=Bq$.
The case $\bar{B}=0$ was considered in the preceding subsection;
henceforth, we therefore assume $\bar{B}\neq0$. In contrast to
Section~\ref{uniform_field_solutions_sec}, however, we now seek
radially dependent solutions, $u=u(r)$, rather than allowing $u$ to
depend on both cylindrical coordinates, $u=u(R,z)$.
\begin{figure*}[t!]
\includegraphics[width=1\textwidth,height=6.5cm,keepaspectratio]{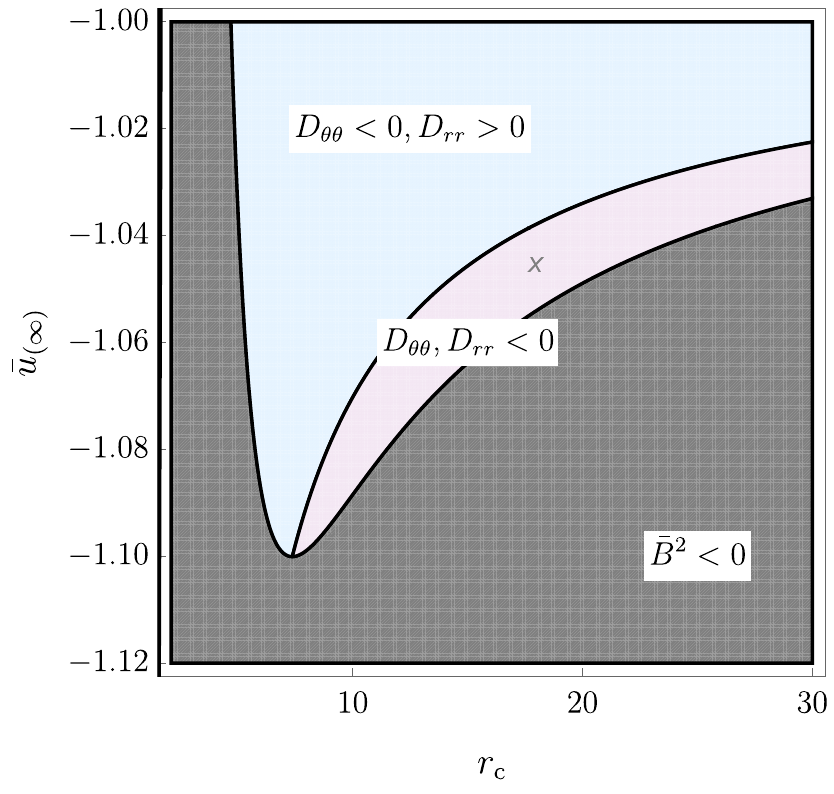}
\includegraphics[width=1\textwidth,height=6.5cm,keepaspectratio]{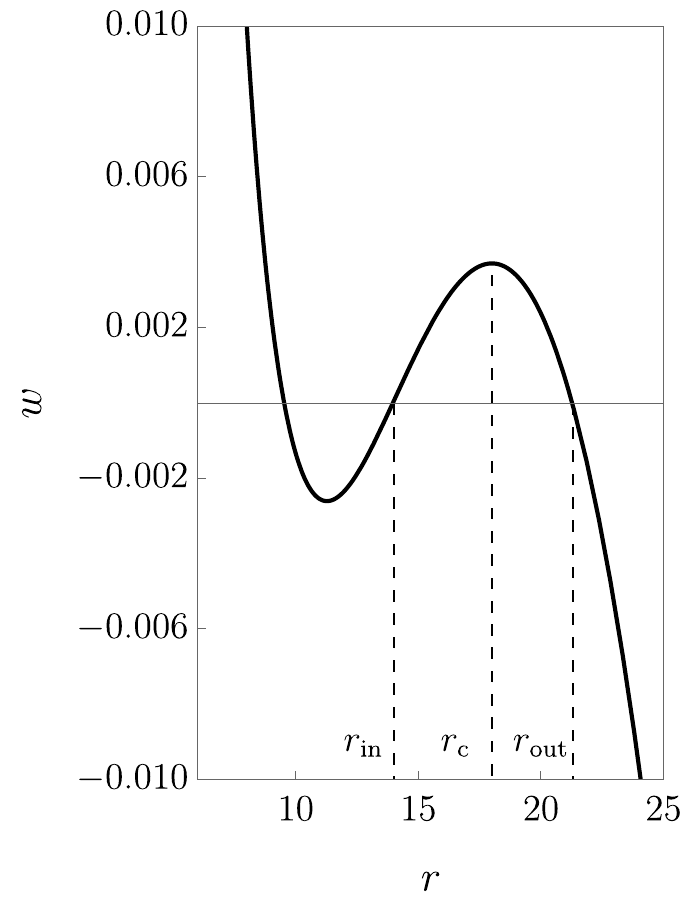}
\includegraphics[width=1\textwidth,height=6.5cm,keepaspectratio]{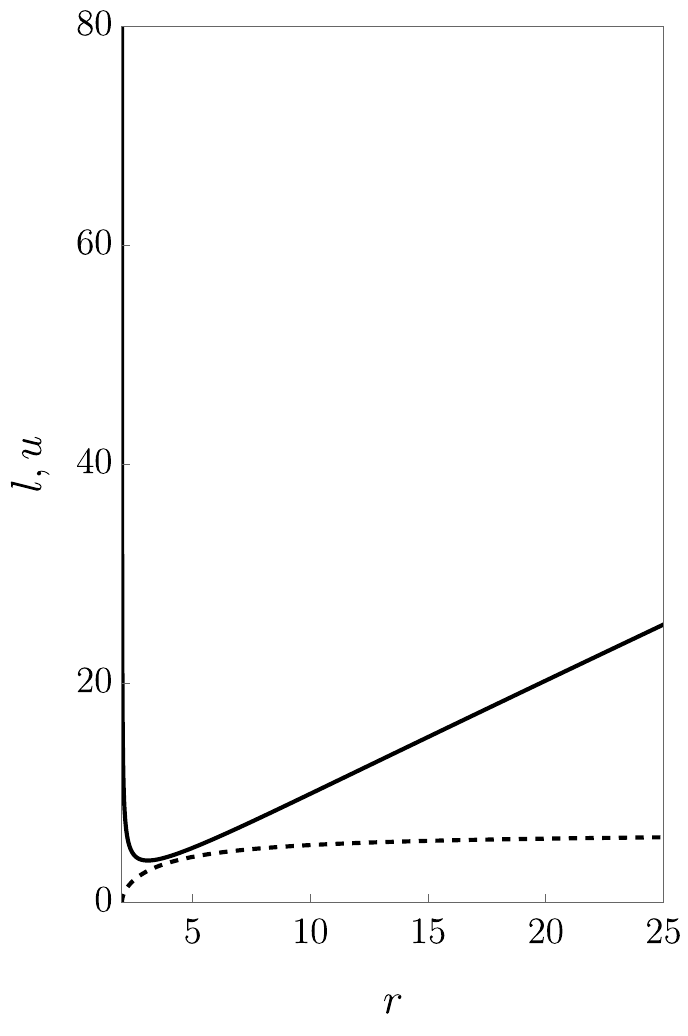}
\includegraphics[width=1\textwidth,height=6.5cm,keepaspectratio]{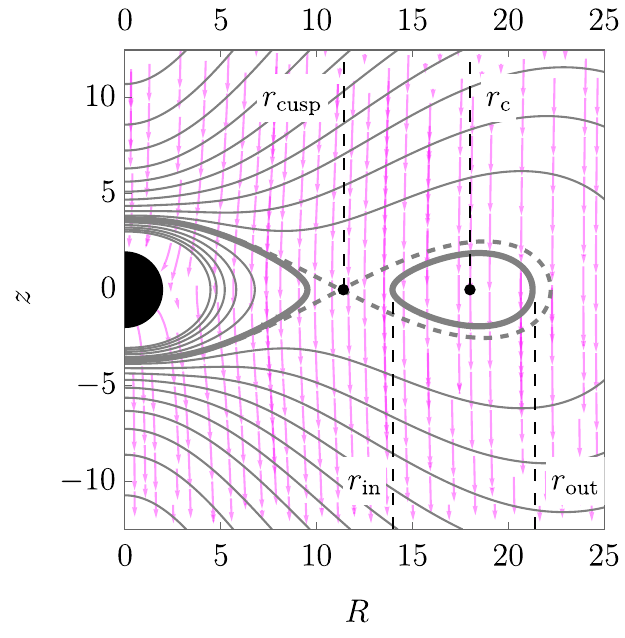}
\begin{minipage}[b]{0.62\textwidth}
   \caption{
Toroidal structures in an asymptotically uniform magnetic field. The
first panel shows the domain of existence of the structures in the
parameter space $(r_{\rm c},\overline{u}_{\infty})$. The purple region
corresponds to candidate centers satisfying both local-maximum
conditions, $D_{\theta\theta}<0$ and $D_{rr}<0$. The blue region,
characterized by $D_{\theta\theta}<0$ and $D_{rr}>0$, corresponds
instead to saddle points of the enthalpy potential $w(R,z)$, whereas the gray region
is inadmissible because the radial stationarity condition would require
$\bar{B}^{\,2}<0$. The cross $(\times)$ marks the parameter choice
$\overline{u}_{\infty}=-1.046$ and $r_{\rm c}=18$, which yields
$\bar{B}=-0.162$ and is used for the representative structure shown
in the remaining panels. The subsequent panels display the equatorial
radial profile $w(r,\pi/2)$ and the corresponding radial profiles of
the parametrization function
$u(r)=\overline{u}(r)/\bar{B}$ and the specific angular momentum
$l(r)$; the integration constant is chosen as
$v_{\mathrm{in}}=-0.3932$. The final panel shows poloidal cross-sections
of the corresponding equipotential surfaces,
$w(R,z)=\mathrm{const}$. The dashed contour is the limiting closed equipotential surface
passing through the cusp, while the thick contour represents a
common closed equipotential surface. The purple curves indicate the magnetic
field lines determined by the orthonormal components $(B^{\hat r},B^{\hat\theta})$.}
\label{Fig:1}
\vspace{10mm}
\end{minipage}
\end{figure*}
For $A(r)=1$, Eq.~\eqref{radial_equation_bar} reduces to
\begin{equation}
    r(r-2)\left(2+\overline{u}\right)\overline{u}'
    =
    2\overline{u}.
\end{equation}
Since $\overline{u}\neq0$, this equation can be separated as
\begin{equation}
    \left(\frac{2}{\overline{u}}+1\right)\mathrm{d}\overline{u}
    =
    \frac{2\,\mathrm{d}r}{r(r-2)}.
\end{equation}
Integration gives
\begin{equation}
    2\ln|\overline{u}|+\overline{u}
    =
    \ln\left(1-\frac{2}{r}\right)+C,
\end{equation}
where $C$ is an integration constant. We parametrize $C$ by the
asymptotic value
\begin{equation}
    \overline{u}_{\infty}
    \equiv
    \lim_{r\to\infty}\overline{u}(r),
\end{equation}
which yields
\begin{equation}
    2\ln\left|\frac{\overline{u}}{\overline{u}_{\infty}}\right|+\overline{u}-\overline{u}_{\infty}
    =
    \ln\left(1-\frac{2}{r}\right).
\label{u_uniform_implicit}
\end{equation}
Exponentiating Eq.~\eqref{u_uniform_implicit}, we obtain
\begin{equation}
    \overline{u}^{\,2}\mathrm{e}^{\overline{u}}
    =
    \overline{u}_{\infty}^{\,2}\mathrm{e}^{\overline{u}_{\infty}}\left(1-\frac{2}{r}\right).
    \label{u_uniform_exponential}
\end{equation}
Since $\overline{u}=\bar{B}u$, this implicit relation can equivalently
be recovered from Eq.~\eqref{F_equation} by choosing
\begin{equation}
    F(x)=\frac{\bar{B}^{\,2}}{\overline{u}_{\infty}^{\,2}\mathrm{e}^{\overline{u}_{\infty}}}x^2.
\end{equation}

Because a nonvanishing solution cannot change sign,
Eq.~\eqref{u_uniform_exponential} can also be written as
\begin{equation}
    \overline{u}\,\mathrm{e}^{\overline{u}/2}
    =
    \overline{u}_{\infty}\,\mathrm{e}^{\overline{u}_{\infty}/2}
    \sqrt{1-\frac{2}{r}}.
\label{u_uniform_lambert_form}
\end{equation}
This relation can be inverted in terms of the Lambert $W$
function.\footnote{Although the Lambert $W$ function cannot generally
be expressed in terms of elementary functions, it is a standard
analytic special function, and the resulting expressions therefore
represent exact analytic solutions.}
The Lambert $W$ function is defined implicitly by
\begin{equation}
    W(x)\mathrm{e}^{W(x)}=x.
\end{equation}
For $\overline{u}_{\infty}\neq0$, the relevant real inversions are
\begin{align}
    \overline{u}_{0}(r)&=2W_{0}\left(\frac{\overline{u}_{\infty}}{2}\mathrm{e}^{\overline{u}_{\infty}/2}
        \sqrt{1-\frac{2}{r}}\right),\;-2\leq\overline{u}_{\infty}<\infty,
\label{eq:u0_expansion}
\\
    \overline{u}_{-1}(r)&=2W_{-1}\left(\frac{\overline{u}_{\infty}}{2}\mathrm{e}^{\overline{u}_{\infty}/2}
        \sqrt{1-\frac{2}{r}}\right),\;-\infty<\overline{u}_{\infty}\leq-2,
\label{eq:u1_expansion}
\end{align}
where $W_0$ and $W_{-1}$ denote the principal and lower real branches,
respectively. At $\overline{u}_{\infty}=-2$, the two solutions have the
same asymptotic value but differ at finite radii.

The radial stationarity condition $D_r=0$, given generally by
Eq.~\eqref{center_parameter_relation}, reduces to
\begin{equation}
    \bar{B}^{\,2}=\frac{\overline{u}_{\rm c}^{\,2}}{r_{\rm c}^{\,2}\left[r_{\rm c}-3+(r_{\rm c}-2)\overline{u}_{\rm c}\right]},
\label{con_drw_B}
\end{equation}
where
$\overline{u}_{\rm c}\equiv\overline{u}(r_{\rm c})$.
Since $\bar{B}\neq0$, its squared value must be strictly positive.
Because the numerator in Eq.~\eqref{con_drw_B} is positive, this
requires
\begin{equation}
    r_{\rm c}-3+(r_{\rm c}-2)\overline{u}_{\rm c}>0
    \quad\Longleftrightarrow\quad
    \overline{u}_{\rm c}>-\frac{r_{\rm c}-3}{r_{\rm c}-2}.
\label{uc_lower}
\end{equation}

To determine whether a candidate center is a strict nondegenerate local
maximum, we must additionally impose the polar and radial conditions
$D_{\theta\theta}<0$ and $D_{rr}<0$, respectively. For $A(r)=1$,
condition~\eqref{d2wtheta} shows that the polar condition is equivalent
to
\begin{equation}
    \overline{u}_{\rm c}>-1.
    \label{uniform_polar_condition}
\end{equation}
However, throughout the exterior region $r_{\rm c}>2$, the lower bound
in Eq.~\eqref{uc_lower} is stronger, because
\begin{equation}
    -\frac{r_{\rm c}-3}{r_{\rm c}-2}=-1+\frac{1}{r_{\rm c}-2}>-1.
\end{equation}
Thus, the polar condition introduces no additional restriction.

Finally, we impose the radial-maximum condition $D_{rr}<0$. For the
asymptotically uniform magnetic field,
\begin{equation}
    A_{\rm c}=1,\qquad A'_{\rm c}=A''_{\rm c}=0.
\end{equation}
Consequently, the quantities entering Eq.~\eqref{Drr_general} reduce
to
\begin{equation}
    \mathcal{K}_{\rm c}=2\left[r_{\rm c}-3+(r_{\rm c}-2)\overline{u}_{\rm c}\right]
\end{equation}
and
\begin{equation}
    \mathcal{K}'_{\rm c}=2+2\overline{u}_{\rm c}+\frac{4\overline{u}_{\rm c}}{r_{\rm c}\left(2+\overline{u}_{\rm c}\right)},
\end{equation}
where Eq.~\eqref{ubar_derivative_center} has been used. Substituting
these expressions into Eq.~\eqref{Drr_general} and eliminating
$\bar{B}^{\,2}$ by means of Eq.~\eqref{con_drw_B}, we obtain
\begin{equation}
    D_{rr}=\frac{\mathcal{P}(r_{\rm c},\overline{u}_{\rm c})}{r_{\rm c}^{\,2}(r_{\rm c}-2)^2(2+\overline{u}_{\rm c})\left[r_{\rm c}-3+(r_{\rm c}-2)\overline{u}_{\rm c}\right]},
\label{Drr_uniform}
\end{equation}
where
\begin{align}
    \mathcal{P}(r_{\rm c},\overline{u}_{\rm c})
    &=(r_{\rm c}-2)(3r_{\rm c}-4)\overline{u}_{\rm c}^{\,2}\nonumber\\
    &\quad+(r_{\rm c}-2)(9r_{\rm c}-16)\overline{u}_{\rm c}\nonumber\\
    &\quad+2\left(3r_{\rm c}^{\,2}-14r_{\rm c}+18\right).
\label{P_uniform}
\end{align}

Under condition~\eqref{uc_lower}, all factors in the denominator of
Eq.~\eqref{Drr_uniform} are positive. Therefore, the sign of $D_{rr}$
is determined entirely by $\mathcal{P}$. Since $\mathcal{P}$ is a
quadratic polynomial in $\overline{u}_{\rm c}$ with a positive leading
coefficient, the condition $D_{rr}<0$ requires
$\overline{u}_{\rm c}$ to lie between its two real roots. The lower
root lies below $-1$ and is excluded by
condition~\eqref{uc_lower}. The remaining restriction is thus
\begin{equation}
   \overline{u}_{\rm c}<\overline{u}_{+}(r_{\rm c}),
\label{uniform_radial_condition}
\end{equation}
where the upper root is
\begin{equation}
    \overline{u}_{+}(r_{\rm c})=\frac{-(9r_{\rm c}-16)+\sqrt{9r_{\rm c}^{\,2}-48-\dfrac{32}{r_{\rm c}-2}}
    }{2(3r_{\rm c}-4)}.
\label{u_plus}
\end{equation}
Consequently, the
local-maximum conditions are satisfied when
\begin{equation}
    -\frac{r_{\rm c}-3}{r_{\rm c}-2}<\overline{u}_{\rm c}<\overline{u}_{+}(r_{\rm c}).
\label{uc_admissible_interval}
\end{equation}

The interval in Eq.~\eqref{uc_admissible_interval} is nonempty only
when
\begin{equation}
    r_{\rm c}>r_*\equiv\frac{1}{2}\left(9+\sqrt{33}\right).
\label{rc_min_uniform}
\end{equation}
At $r_{\rm c}=r_*$, its lower and upper bounds coincide.

On the principal branch, the function
$\overline{u}\mathrm{e}^{\overline{u}/2}$ is strictly increasing
throughout the relevant domain $\overline{u}>-1$. Therefore,
Eq.~\eqref{u_uniform_lambert_form} maps the lower and upper limits in
Eq.~\eqref{uc_admissible_interval} to
\begin{equation}
    \overline{u}_{\infty}>2W_0\left(-\frac{(r_{\rm c}-3)\sqrt{r_{\rm c}}}{2(r_{\rm c}-2)^{3/2}}
        \mathrm{e}^{-\frac{r_{\rm c}-3}{2(r_{\rm c}-2)}}\right)
\label{urc_lower}
\end{equation}
and
\begin{equation}
    \overline{u}_{\infty}<2W_0\left(\frac{\overline{u}_{+}(r_{\rm c})}{2}\mathrm{e}^{\overline{u}_{+}(r_{\rm c})/2}\frac{1}{\sqrt{1-\dfrac{2}{r_{\rm c}}}}\right),
\label{urc_upper}
\end{equation}
respectively. Thus, for each fixed $r_{\rm c}>r_*$, the asymptotic
parameter $\overline{u}_{\infty}$ must lie strictly between the
radius-dependent bounds in Eqs.~\eqref{urc_lower} and
\eqref{urc_upper}.

At $r_{\rm c}=r_*$, the right-hand side of
Eq.~\eqref{urc_lower} attains its absolute minimum. The projections of
the admissible parameter domain onto the
$\overline{u}_{\infty}$ and $r_{\rm c}$ axes are therefore
\begin{align}
    2W_0\left(-\frac{(3+\sqrt{33})\sqrt{9+\sqrt{33}}}{2(5+\sqrt{33})^{3/2}}\mathrm{e}^{-\frac{1}{8}(9-\sqrt{33})}\right)<\overline{u}_{\infty}<-1,\\
    \frac{1}{2}\left(9+\sqrt{33}\right)<r_{\rm c}<\infty.
\end{align}
Numerically, $-1.10007\ldots<\overline{u}_{\infty}<-1$, and $7.37228\ldots<r_{\rm c}<\infty$.
These overall ranges are projections of the admissible domain and
should not be interpreted as independent sufficient conditions. For
each fixed $r_{\rm c}$, the radius-dependent bounds in
Eqs.~\eqref{urc_lower} and \eqref{urc_upper} must still be satisfied.

A representative configuration with
$\overline{u}_{\infty}=-1.046$ and $r_{\rm c}=18$ is shown in
Fig.~\ref{Fig:1}. Values within the overall interval
$-1.10007\ldots<\overline{u}_{\infty}<-1$ can support 
toroidal structures associated with a local maximum of $w$,
provided that the radius-dependent bounds in
Eqs.~\eqref{urc_lower} and \eqref{urc_upper} are also satisfied.
For a fixed solution $\overline{u}(r)$, the relation
$u=\overline{u}/\bar{B}$ implies that the sign of the angular momentum
is determined by the sign of $\bar{B}$. Consequently, changing the sign
of $\bar{B}$ reverses the sense of rotation while leaving the enthalpy
potential unchanged.

On the lower Lambert branch,
$\overline{u}=2W_{-1}\leq-2$. This is incompatible with the
polar-maximum condition $\overline{u}_{\rm c}>-1$. Consequently, this
branch cannot admit an equatorially centered local maximum of $w$ and
therefore does not yield the toroidal structures considered here.

\subsection{Combined magnetic fields}
We now consider the superposition of the asymptotically uniform
external magnetic field and the regularized dipole approximation to
the magnetic self-field introduced above. For $B\neq0$, we set
\begin{equation}
A(r)=1+\frac{2\tilde{I}}{\left(r^2+r_0^2\right)^{3/2}},
\qquad
\mu=B,\qquad\tilde{I}\equiv\frac{I}{B},
\label{total_A_loop}
\end{equation}
where $\tilde{I}$ measures the effective self-field amplitude
relative to the strength of the external field. Thus, the corresponding
azimuthal vector potential is
\begin{equation}
\mathcal{A}_{\phi}=\frac{B}{2}r^2\sin^2\theta+\frac{I r^2\sin^2\theta}{\left(r^2+r_0^2\right)^{3/2}}.
\end{equation}
\begin{figure*}[t!]
\includegraphics[width=1\textwidth,height=6.5cm,keepaspectratio]{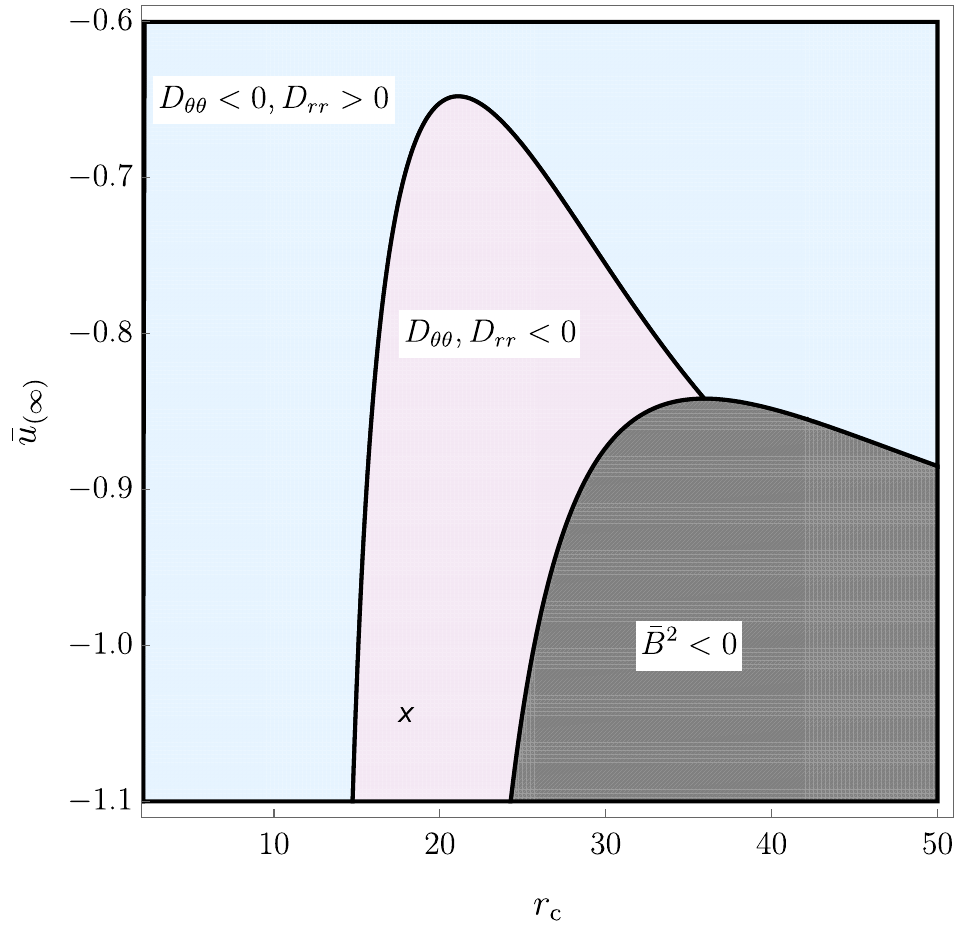}
\includegraphics[width=1\textwidth,height=6.5cm,keepaspectratio]{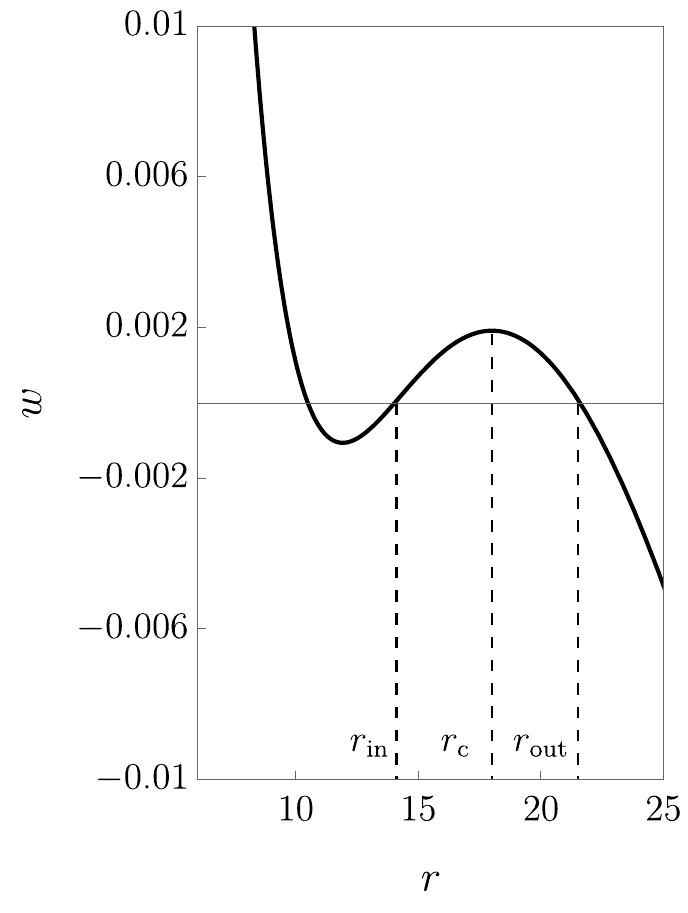}
\includegraphics[width=1\textwidth,height=6.5cm,keepaspectratio]{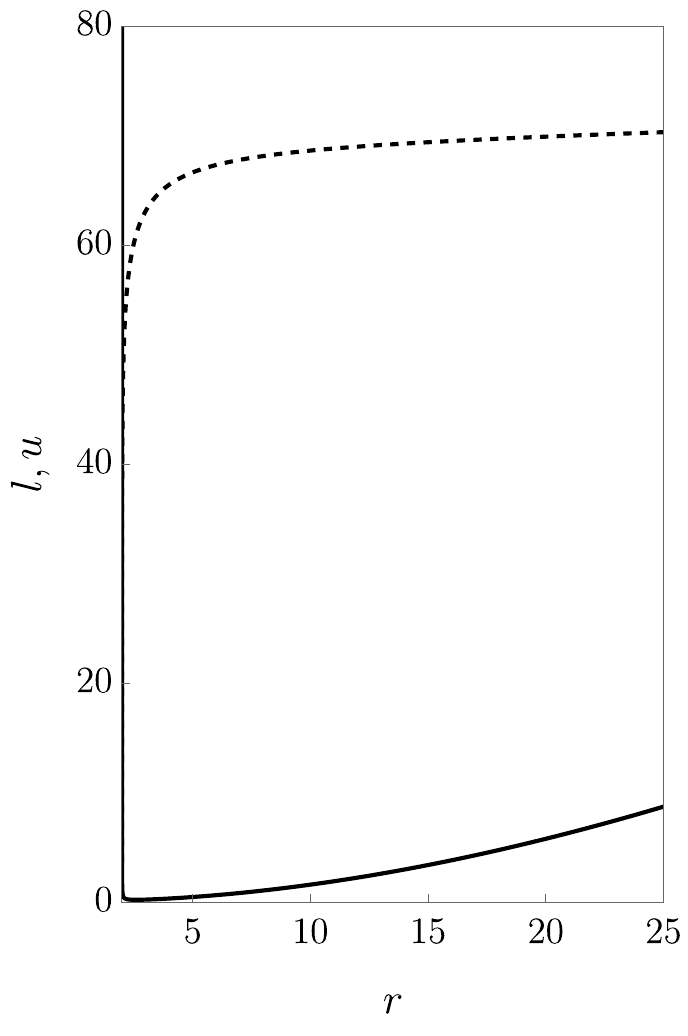}
\includegraphics[width=1\textwidth,height=6.5cm,keepaspectratio]{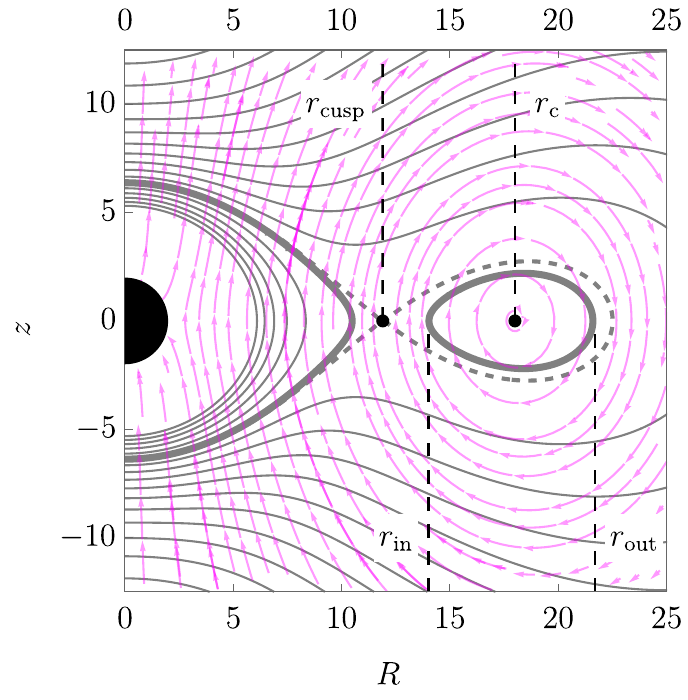}\hfill
\begin{minipage}[b]{0.62\textwidth}
\caption{Toroidal structures in a combined magnetic-field configuration
consisting of an asymptotically uniform background field and an
approximate current-loop self-field. The first panel shows the domain of existence in parameter space $(r_{\rm c},\overline{u}_{\infty})$. The purple region corresponds to candidate centers satisfying both local-maximum
conditions, $D_{\theta\theta}<0$ and $D_{rr}<0$. The blue region,
characterized by $D_{\theta\theta}<0$ and $D_{rr}>0$, corresponds
instead to saddle points of the enthalpy potential $w(R,z)$, whereas the gray region
is inadmissible because the radial stationarity condition would require
$\bar{B}^{\,2}<0$.
The cross $(\times)$ marks the parameter choice $\overline{u}_{\infty}=-1.046$, $r_{0}=r_{\rm c}=18$, and $\tilde{I}=-4\sqrt{2}r_{0}^3$, yielding $\bar{B}=0.014$. Remaining panels display equatorial profiles of $w(r,\pi/2)$, $u(r)=\overline{u}(r)/\bar{B}$, and $l(r)$ for $v_{\mathrm{in}}=-0.1881$, and poloidal cross-sections of the equipotential surfaces $w(R,z)=\mathrm{const}$. The dashed contour is the limiting closed equipotential surface
passing through the cusp, while the thick contour represents a
common closed equipotential surface. The purple curves indicate the magnetic
field lines determined by the orthonormal components
$(B^{\hat r},B^{\hat\theta})$.
}
\label{Fig:2}
\vspace{15mm}
\end{minipage}
\end{figure*}
To obtain asymptotic data for the numerical integration of
Eq.~\eqref{radial_equation_bar}, we expand the radial solution as
\begin{equation}
\overline{u}(r)
=
\sum_{n=0}^{\infty}
\frac{a_n}{r^n}.
\label{combined_u_expansion}
\end{equation}
Substitution of Eq.~\eqref{combined_u_expansion} into
Eq.~\eqref{radial_equation_bar} and matching equal powers of $1/r$
gives
\begin{eqnarray*}
    a_{0}&=&\overline{u}_{\infty},\quad
    a_{1}=-\frac{2\overline{u}_{\infty}}{2+\overline{u}_{\infty}},\quad
    a_{2}=-\frac{2\overline{u}_{\infty}(2+4\overline{u}_{\infty}+\overline{u}_{\infty}^2)}{(2+\overline{u}_{\infty})^3},\\
    a_{3}&=&-\frac{8\overline{u}_{\infty}(6+\overline{u}_{\infty}(2+\overline{u}_{\infty})(3+\overline{u}_{\infty})^2)}{3(2+\overline{u}_{\infty})^5},\\
        a_{4}&=&-\frac{4\overline{u}_{\infty}}{3(2+\overline{u}_{\infty})^7}\Big(60+240\overline{u}_{\infty}+414\overline{u}_{\infty}^2+380\overline{u}_{\infty}^3\Big.\\
        &+&\left.169\overline{u}_{\infty}^4+36\overline{u}_{\infty}^5+3\overline{u}_{\infty}^6\right)+\frac{\overline{u}_{\infty}^2\tilde{I}}{(2+\overline{u}_{\infty})^2}.
\end{eqnarray*}
The contribution of the magnetic self-field first enters through the
term proportional to $\tilde{I}$ in $a_4$ and is therefore of
order $\mathcal{O}(r^{-4})$. It is consequently subleading in the
asymptotic region. The expansion truncated at a sufficiently high
order provides initial data at a large but finite radius for the
numerical integration toward the torus.

To define a fiducial relative self-field amplitude, we impose the
additional modeling condition
\begin{equation}
B^{\hat{\theta}}(r_0,\pi/2)=0.
\label{vertical_field_condition}
\end{equation}
According to Eq.~\eqref{measured_magnetic_field}, this is equivalent,
for $r_0>2$, to
\begin{equation}
\left.
\partial_r\mathcal{A}_{\phi}
\right|_{(r_0,\pi/2)}
=
0.
\end{equation}
For the combined vector potential,
\begin{equation}
\left.\partial_r\mathcal{A}_{\phi}\right|_{\theta=\pi/2}=Br\left(1+\tilde{I}\frac{2r_0^2-r^2}{\left(
r^2+r_0^2\right)^{5/2}}\right).
\end{equation}
Evaluating this expression at $r=r_0$ and imposing
Eq.~\eqref{vertical_field_condition} yields
\begin{equation}
    \tilde{I}=\tilde{I}_0\equiv-4\sqrt{2}\,r_0^3.
\label{current_amplitude_condition}
\end{equation}
We refer to $\tilde{I}_0$ as the fiducial relative self-field
amplitude. The magnetic-null configuration corresponds to
$\tilde{I}=\tilde{I}_0$, while the dependence on the self-field
strength is explored by setting
$\tilde{I}=k\tilde{I}_0$.
\begin{figure*}[t!]
\centering 
\includegraphics[width=1\textwidth,height=5.25cm,keepaspectratio]{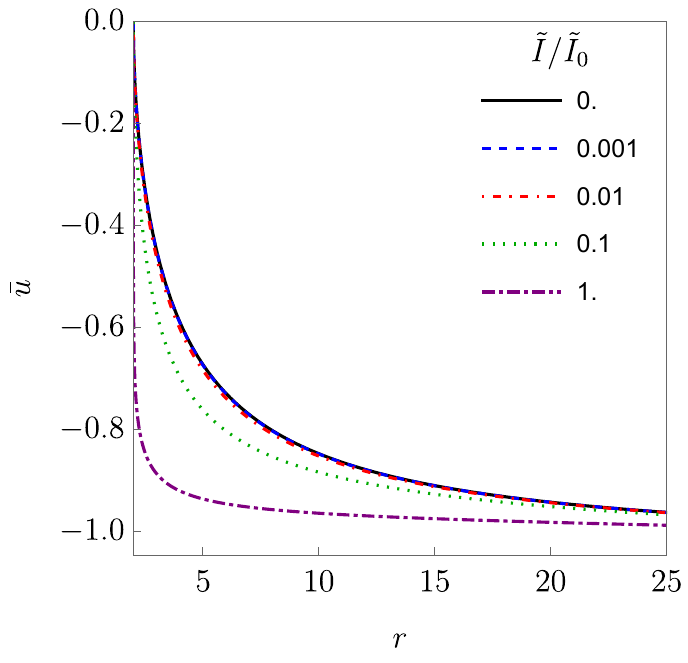}
\includegraphics[width=1\textwidth,height=5.25cm,keepaspectratio]{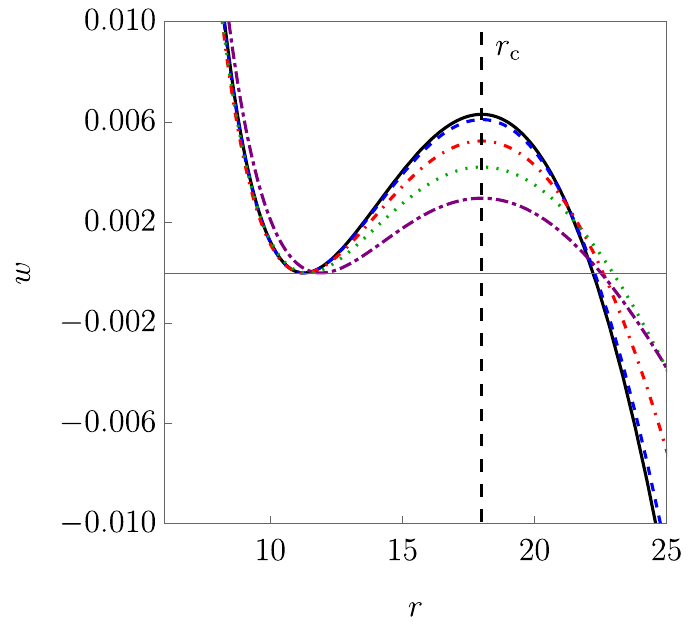}
\includegraphics[width=1\textwidth,height=5.25cm,keepaspectratio]{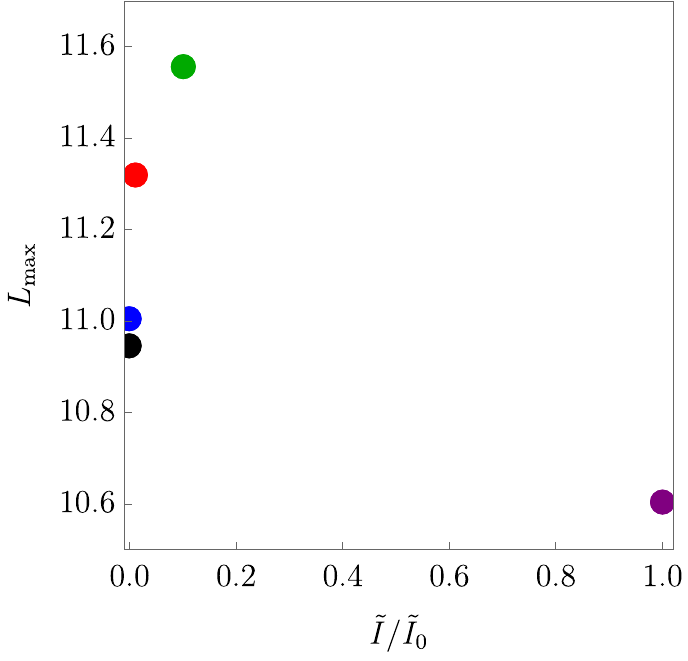}
\caption{Effect of the approximate self-field amplitude on toroidal structures
in the combined magnetic configuration. The first panel shows radial
profiles of the rescaled angular-momentum parametrization function
$\overline{u}(r)$, the middle panel shows the equatorial enthalpy
profiles $w(r,\pi/2)$, and the final panel shows the radial extent
$L=r_{\rm out}-r_{\rm cusp}$ as a function of
$\tilde{I}/\tilde{I}_0$. The profiles correspond to
$k=0.001$, $0.01$, $0.1$, and $1$, with $r_0=18$.}
\label{Fig:3}
\end{figure*}

Since the radial magnetic-field component vanishes on the equatorial
plane by symmetry, $(r_0,\pi/2)$ is a magnetic null. At this point, the
self-generated and external magnetic fields have equal magnitudes and
opposite directions. The equatorial vertical component changes sign
across the null: it has the polarity opposite to that of the external
field for $2<r<r_0$, while for $r>r_0$ it has the same polarity and
approaches the asymptotically uniform field at large radii.

For the representative combined-field configuration, we set $r_0=r_{\rm c}$, thereby identifying the effective loop radius with the
location of the pressure maximum and hence with the density core of the
torus. This phenomenological choice is motivated by the expectation
that a sufficiently slender torus can be approximated by a circular
current loop centered at $r_{\rm c}$. It does not imply that the
regularized dipole profile coincides with the exact field of a thin
current loop.

The parameters $\tilde{I}$, $r_0$, and $\overline{u}_{\infty}$ determine the radial solution $\overline{u}(r)$, whereas $\bar{B}=Bq$ and $v_{\mathrm{in}}$ enter the corresponding enthalpy potential. Together, these five parameters specify a particular toroidal structure. A representative numerical solution with $\tilde{I}=-4\sqrt{2}r_0^3\equiv\tilde{I}_0$, $r_0=18$, $\overline{u}_{\infty}=-1.046$ 
is shown in Fig.~\ref{Fig:2}. The remaining parameter values used to construct this structure are specified in the figure caption. Near $r=r_0$, the self-field is comparable to the external field and can therefore significantly modify the Lorentz-force balance and the resulting toroidal structure. Figure~\ref{Fig:3} shows that the radial extent $L=r_{\rm out}-r_{\rm cusp}$ depends nonmonotonically on the relative
self-field amplitude. It increases for the weak self-fields considered
here, $k\ll1$, but decreases again as $k$ approaches unity. This trend is consistent with the self-field-only results presented in the next subsection, where no toroidal structure is found within the
parameter domain examined numerically.
\subsection{Approximate current-loop self-field}\label{current_loop_sec}
We finally consider the regularized dipole approximation to the
magnetic self-field separately, with the external field omitted. In
this case, we set
\begin{equation}
A(r)=\frac{2}{\left(r^2+r_0^2\right)^{3/2}},\qquad\mu=I,
\label{current_loop_field}
\end{equation}
so that $\bar{\mu}=\bar{I}\equiv Iq$.

To obtain asymptotic data for the numerical integration, we expand the
rescaled parametrization function as
\begin{equation}
\overline{u}(r)=\sum_{n=0}^{\infty}\frac{a_n}{r^n}.
\label{loop_u_expansion}
\end{equation}
Substitution of Eq.~\eqref{loop_u_expansion} into
Eq.~\eqref{radial_equation_bar} gives
\begin{eqnarray*}
    a_{0}&=&\overline{u}_{\infty},\quad
    a_{1}=-\overline{u}_{\infty},\quad
    a_{2}=-\frac{\overline{u}_{\infty}}{2},\quad
    a_{3}=-\frac{\overline{u}_{\infty}}{2},\\
    a_{4}&=&\frac{1}{8}\left(2\overline{u}_{\infty}^2-5\overline{u}_{\infty}\right),\quad
    a_{5}=-\frac{1}{40}\left(2\overline{u}_{\infty}^2+35\overline{u}_{\infty}\right),\\
        a_{6}&=&-\frac{1}{80}\left(6\overline{u}_{\infty}^2+20\overline{u}_{\infty}^2r_{0}^2+105\overline{u}_{\infty}\right),\\
\end{eqnarray*}
\begin{figure*}[t!]
\centering 
\includegraphics[width=1\textwidth,height=5.25cm,keepaspectratio]{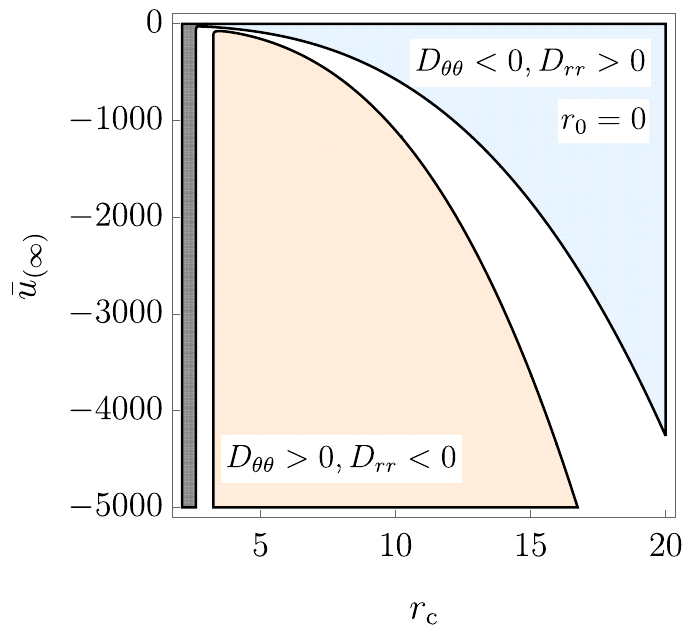}
\includegraphics[width=1\textwidth,height=5.25cm,keepaspectratio]{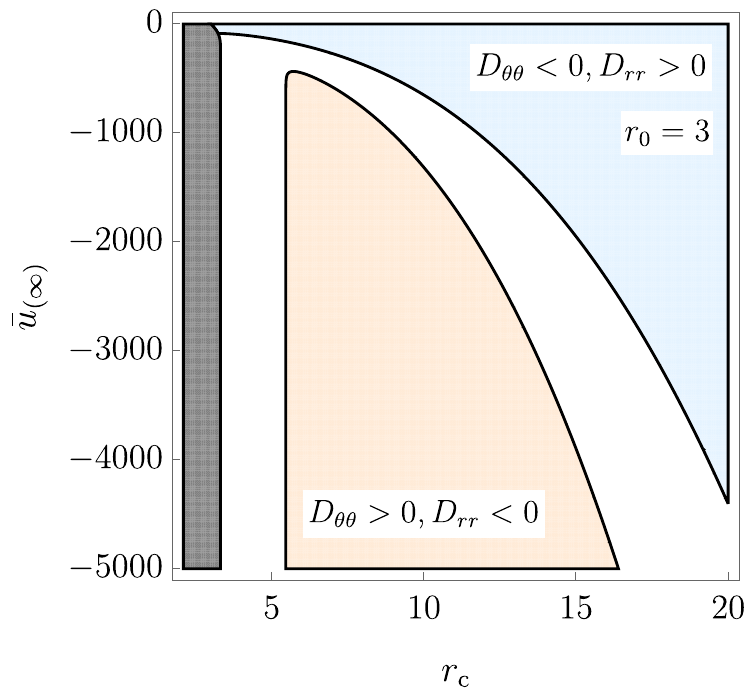}
\includegraphics[width=1\textwidth,height=5.25cm,keepaspectratio]{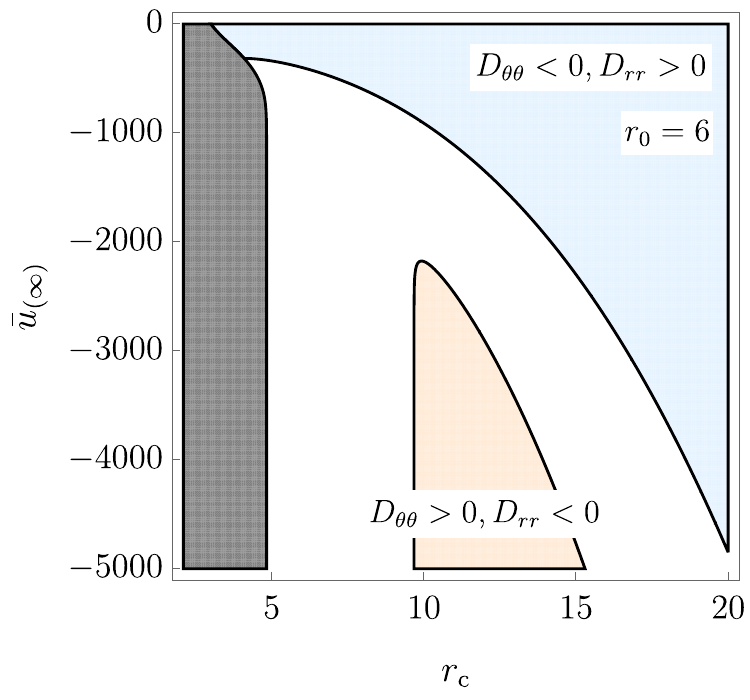}

\caption{Absence of toroidal structures in the numerically
examined parameter domains for the isolated approximate current-loop
self-field. The panels show the parameter space
$(r_{\rm c},\overline{u}_{\infty})$ for three regularization scales:
$r_0=0$ (left), $r_0=3$ (middle), and $r_0=6$ (right). The gray regions
indicate the inadmissible domain $\bar{I}^{\,2}<0$. The light-blue
regions satisfy $D_{\theta\theta}<0$ and $D_{rr}>0$, while the
light-orange regions satisfy $D_{rr}<0$ and
$D_{\theta\theta}>0$. In the white regions, both second derivatives
are positive. Since $D_{\theta\theta}<0$ and $D_{rr}<0$ are nowhere
satisfied simultaneously, no equatorially centered toroidal
structure is found within the displayed parameter domains.}
\label{Fig:4}
\end{figure*}
Figure~\ref{Fig:4} shows the parameter space
$(r_{\rm c},\overline{u}_{\infty})$ for the regularization scales
$r_0=0$, $3$, and $6$. The candidate center radii $r_{\rm c}$ are
determined by the radial stationarity condition $D_r=0$, after which
the polar and radial local-maximum conditions are evaluated.

For $r_0=0$, the adopted profile reduces to
\begin{equation}
A(r)=\frac{2}{r^3},
\qquad
A_{\phi}^{\mathrm{self}}
=
\frac{I\sin^2\theta}{r}.
\end{equation}
This is the unregularized point-dipole limit of the present
approximation rather than the exact field of a circular current loop.
It is nevertheless admissible throughout the exterior Schwarzschild
region $r>2$, where it remains regular; its singularity occurs only at
$r=0$, outside the domain considered here.

Within the parameter ranges examined numerically, the two conditions
$D_{\theta\theta}<0$ and $D_{rr}<0$ are never satisfied simultaneously for any of the three selected
values of $r_0$. Thus, no strict nondegenerate local maximum of $w$ is
found, and consequently no equatorially centered closed toroidal
structure is obtained within the investigated part of this
self-field-only class.
\section{Conclusions}
\label{discussion_conclusion_sec}

In this paper, we developed an analytic and semi-analytic framework for
modeling stationary equilibrium structures of a charged,
non-self-gravitating perfect fluid orbiting a Schwarzschild black hole
in prescribed poloidal magnetic-field configurations. We considered
the zero-conductivity limit and assumed a cold, rest-mass-dominated
fluid with a spatially uniform specific charge. Throughout the
existence analysis, we restricted our attention to configurations that
are reflection-symmetric with respect to the equatorial plane and whose
centers lie in that plane. Thus, the existence results presented here concern equatorially
centered tori only.

Under the adopted assumptions, the integrability condition for the
relativistic Euler equations was reduced to a first-order quasilinear
partial differential equation for the angular-momentum parametrization
function $u$. For an asymptotically uniform external magnetic field
with the magnetic self-field neglected, the method of characteristics
yielded a general implicit integral for the class $u=u(R,z)$. We also recovered separately the cylindrically dependent class $u=u(R)$, previously studied, for example, in
Ref.~\cite{2020CQGra..37x5007K}. For this class,
the angular velocity and specific angular momentum satisfy
$\Omega l=k$, and hence $\Omega=k/l$, providing an explicit
nonconstant rotation law of the form $\Omega=\Omega(l)$.

In the field-free and uncharged-fluid limits, the general integrability
condition reduces to the standard integrability structure of an
uncharged barotropic perfect fluid. For nonconstant
specific-angular-momentum profiles, this structure is expressed by the
functional relation $\Omega=\Omega(l)$, whereas constant-$l$ profiles
form a degenerate, automatically integrable subclass. The latter
includes the well-known constant-specific-angular-momentum tori \cite{Abramowicz1978,Kozlowski1978}.

We then turned to the radially dependent class $u=u(r)$. For this
class, we integrated the Euler equations, obtained the corresponding
enthalpy potential, and derived the equatorial stationarity and
local-maximum conditions for a general radial factor $A(r)$ entering
the adopted separable ansatz for the azimuthal vector potential of a
poloidal magnetic field. In the field-free and uncharged-fluid limits,
however, the candidate centers within this restricted class correspond
to radial minima of the enthalpy potential. Thus, the radially
dependent class does not admit equatorially centered closed toroidal
equilibrium structures in these limits. This negative result is
specific to the class $u=u(r)$ and therefore does not contradict the
existence of the standard constant-$l$ tori recovered within the more
general integrability framework.

When an asymptotically uniform external magnetic field is introduced,
the radial integrability equation can instead be solved exactly in
terms of the Lambert $W$ function. Its principal branch admits
equatorially centered local maxima within the derived radius-dependent
parameter bounds and therefore allows the corresponding closed
toroidal structures to be constructed. In contrast, the lower real
branch is excluded because it is incompatible with the polar-maximum
condition. Moreover, for a fixed rescaled parametrization function,
changing the sign of the effective magnetic parameter reverses the
sense of rotation while leaving the enthalpy potential unchanged.

To represent the magnetic self-field of the torus, we adopted a
regularized dipole approximation to the field of an effective circular
current loop. When this approximate self-field is superposed on the
asymptotically uniform external field, the relative loop amplitude can
be chosen so that the vertical magnetic-field component vanishes at
the effective loop radius, thereby producing an equatorial magnetic
null. The representative combined-field solution then shows that the
resulting local cancellation between the two magnetic components can
substantially modify the enthalpy distribution and the morphology of
the torus. Moreover, for the representative sequence examined, the
radial extent $L=r_{\rm out}-r_{\rm cusp}$ depends nonmonotonically on
the relative self-field amplitude: it initially increases for weak
self-fields but decreases again as the approximate self-field becomes
comparable to the external field. Thus, the self-field does not merely
expand or compress the torus. Rather, its influence results from the
changing local balance between the external and torus-generated
magnetic components.

In contrast, for the isolated approximate current-loop self-field, the
polar and radial local-maximum conditions were not satisfied
simultaneously within the parameter ranges and loop radii examined
numerically. Consequently, no equatorially centered closed toroidal
equilibrium was found in the investigated part of this self-field-only
class. However, this numerical result should not be interpreted as a
general nonexistence proof outside the explored parameter domain. Taken
together, the results indicate that, within the configurations examined
here, the asymptotically uniform external component provides the
magnetic background on which the approximate self-field can
significantly reshape an otherwise admissible toroidal equilibrium.

\begin{table}[htbp]
\centering
\caption{
Existence of toroidal structures within the radially dependent class
$u=u(r)$. Numerical results refer to the parameter domains examined.}
\label{tab:torus_structure}
\small
\begin{tabularx}{\linewidth}{>{\raggedright\arraybackslash}X>{\raggedright\arraybackslash}X}
\toprule
\textbf{Magnetic field configuration}&\textbf{Toroidal structures}
\\
\midrule
Field-free or uncharged-fluid limit&No (analytic)
\\
\addlinespace
Asymptotically uniform external field&Yes (analytic bounds)
\\
\addlinespace
Approximate current-loop self-field only&Not found (numerical)
\\
\addlinespace
Combined external and approximate self-fields&Found (numerical)
\\
\bottomrule
\end{tabularx}
\end{table}
The principal existence results for closed structures within the
radially dependent class are summarized in
Table~\ref{tab:torus_structure}.
It should be emphasized, however, that
the effective current-loop model remains phenomenological: its
amplitude is prescribed rather than determined self-consistently from
the distributed azimuthal current of the fluid. In addition, both
magnetic components are treated as test fields with respect to the
background spacetime, while the electrostatic self-field is neglected.
The conclusions are also tied to the cold-fluid, zero-conductivity, and
uniform-specific-charge assumptions. Nevertheless, the resulting
framework clarifies how rotation, pressure, and an approximate magnetic
self-field enter the equilibrium conditions and provides a useful basis
for more self-consistent models. Natural extensions therefore include
off-equatorial configurations, rotating black-hole backgrounds, finite
electrical conductivity, electrostatic self-fields, and a
self-consistent coupling between the fluid current and Maxwell's
equations.
\begin{acknowledgements}
The authors acknowledge the institutional support of the Research Centre for Theoretical Physics and Astrophysics, Institute of Physics, Silesian University in Opava. The work was also supported by the internal grants IGS/27/2026 and SGS/24/2024 of the Silesian University in Opava. AT acknowledges  the support of science and research in the Moravian-Silesian Region 2023 Grant No.  RRC/09/2023.
\end{acknowledgements}
\bibliographystyle{aasjournalv7}
\bibliography{references}
\end{document}